\documentclass[11pt,letterpaper]{article}
\usepackage[margin=1in]{geometry}
\usepackage[T1]{fontenc}
\usepackage[utf8]{inputenc}
\usepackage{lmodern}
\usepackage{amsmath,amssymb,textcomp}
\usepackage{graphicx}
\usepackage{booktabs,array,tabularx}
\usepackage{caption,float,placeins,pdflscape}
\usepackage{microtype}
\usepackage{cite}
\usepackage[hidelinks]{hyperref}
\DeclareUnicodeCharacter{2265}{\ensuremath{\geq}}
\DeclareUnicodeCharacter{2264}{\ensuremath{\leq}}
\DeclareUnicodeCharacter{2212}{\ensuremath{-}}
\DeclareUnicodeCharacter{22A5}{\ensuremath{\perp}}
\DeclareUnicodeCharacter{0394}{\ensuremath{\Delta}}
\DeclareUnicodeCharacter{03C0}{\ensuremath{\pi}}
\DeclareUnicodeCharacter{2003}{\quad}
\DeclareUnicodeCharacter{2009}{\,}
\DeclareUnicodeCharacter{2010}{-}
\hypersetup{pdftitle={TorchSpin: Differentiable Electron Paramagnetic Resonance Simulation in PyTorch},pdfauthor={Alec H. Follmer}}

\begin{document}

\begin{center}
{\LARGE\bfseries TorchSpin: Differentiable Electron Paramagnetic Resonance
Simulation in PyTorch\par}
\vspace{12pt}
Alec H. Follmer\textsuperscript{1,2*}\par
\vspace{6pt}
{\small \emph{\textsuperscript{1} Department of Chemistry, University of
California, Davis, Davis, California 95616, United States}\par}
{\small \emph{\textsuperscript{2} Linac Coherent Light Source, SLAC National
Accelerator Laboratory, Menlo Park, California 94025, United States}\par}
{\small \textsuperscript{*}Corresponding author:
\href{mailto:ahfollmer@ucdavis.edu}{\nolinkurl{ahfollmer@ucdavis.edu}}\par}
\end{center}

\begin{abstract}
Interpreting electron paramagnetic resonance (EPR) spectra commonly
requires repeated forward simulation and optimization of spin
Hamiltonian parameters. Established software provides mature simulation
and fitting methods, but these calculations are not generally available
as differentiable components within machine-learning workflows.
We present TorchSpin, an open-source PyTorch framework that ports the
organization and numerical conventions of EasySpin into a tensor-based
computational environment. Representative calculations spanning
continuous-wave EPR, ENDOR, pulse EPR, spin dynamics, trajectory
simulation, and magnetometry closely reproduced corresponding EasySpin
results. In matched benchmarks, TorchSpin CPU execution was faster than
EasySpin for six of nine workloads although performance depended
strongly on the calculation. We therefore focused on the capabilities
enabled by automatic differentiation in PyTorch. In a Cu(II)
powder-spectrum demonstration, feedback propagated through TorchSpin
improved the quality of a supervised convolutional neural network (CNN)
without using parameter labels during refinement. Models trained from
randomly initialized CNNs using only noisy spectra and TorchSpin
reconstruction loss achieved median similarities of ≥ 0.99. CNNs trained on TorchSpin-generated libraries also
transferred without adaptation to independently generated EasySpin
spectra with performance comparable to EasySpin-trained controls. Once trained, the CNN produced initial parameter estimates in approximately 1 ms per spectrum while a batch of 512 was processed in approximately 1.3 ms with GPU acceleration. TorchSpin therefore provides both a tested Python
implementation of established EPR calculations and a differentiable
physical layer for rapid learned inference, spectral refinement, and
parameter-label-free training.
\end{abstract}


\FloatBarrier
\section*{Introduction}

Quantitative interpretation of electron paramagnetic resonance (EPR)
spectra usually depends on forward simulation.\cite{ref1,ref2,ref3,ref4,ref5} Except in the
simplest cases, the principal values and orientations of the \emph{g},
hyperfine, quadrupole, and zero-field-splitting tensors cannot be read
directly from the experimental signal. Instead, a spin Hamiltonian is
constructed, its response is evaluated over the experimental field,
frequency, or time coordinates, and the simulated spectrum is compared
with the measurement. The accuracy and computational structure of the
forward model therefore determine which spin systems can be analyzed and
how their parameters can be estimated. The performance of nonlinear
regression algorithms dictates the speed and complexity of parameter
refinement.

Established simulation packages have made this approach routine across
much of EPR spectroscopy. EasySpin provides particularly broad and
extensively validated methods for continuous-wave EPR, ENDOR, pulse EPR,
spin dynamics, magnetometry, and spectral fitting, \cite{ref1,ref2} while
Spinach \cite{ref4}, XSophe \cite{ref5}, and more specialized programs provide
complementary treatments of spin dynamics and particular experimental
regimes. Together, these packages establish a high standard for physical
coverage and numerical reliability. Machine-learning approaches have
also been developed for specific EPR inverse problems, including the
extraction of magnetic coupling parameters from HYSCORE spectra\cite{ref6}
and distance distributions from pulsed dipolar measurements \cite{ref7,ref8}.
These studies demonstrate how networks trained on simulated data can
support experimental analysis within defined measurement regimes. The
computational interfaces of general-purpose simulation packages,
however, were developed primarily for conventional forward simulation
and iterative optimization. Connecting these calculations to
machine-learning systems commonly requires spectra and parameters to be
transferred between separate programs, and the dependence of the
simulated signal on the underlying parameters is not generally available
through automatic differentiation.

This distinction matters because learned spectral analysis can use a
simulator in several different ways. A conventional supervised model can
be trained on a library of simulated spectra and the
parameters that generated them, after which it can provide a rapid initial estimate for each
new measurement. In this setting, the simulator acts only as a source of
labeled training data and does not need to be differentiable. A
differentiable simulator enables a second form of training: parameters
predicted from a spectrum can be returned to the physical model, and
disagreement between the reconstructed and measured spectra can be
propagated through the simulation to update either the predicted
parameters or the network itself \cite{ref9,ref10,ref11}. The spectrum can therefore
provide a training signal even when its generating Hamiltonian is
unknown. More generally, measured spectra, learned models, and the
physical simulation can participate in the same computation, allowing
physical constraints to be incorporated directly into optimization
rather than applied only after a prediction has been made.

Here we present TorchSpin, an open-source PyTorch framework \cite{ref12}
that ports the organization, program names, and numerical conventions of
EasySpin into a differentiable tensor-based environment. Rather than
attempting to catalog every available option, we focus on the
requirements that determine whether the framework can support scientific
analysis: numerical correspondence across representative EPR regimes,
performance during individual and repeated simulations, reliable
execution on CPU and accelerator hardware, and the ability to connect a
forward calculation directly to learned parameter inference. We validate
eight TorchSpin simulation pathways against stored EasySpin calculations
and benchmark nine workloads on matched hardware. We then use noisy
Cu(II) powder spectra to examine three complementary applications:
differentiable refinement of an already trained inverse model,
parameter-label-free training from random initialization, and supervised
transfer from TorchSpin-generated libraries to independently generated
EasySpin spectra. These tests define the current validated scope of
TorchSpin while illustrating how established EPR physics can be used as
both a data generator and a differentiable decoder.

\FloatBarrier
\section*{Results}

\subsection*{TorchSpin reproduces EasySpin across representative simulation regimes}

TorchSpin is a PyTorch port of the EasySpin simulation framework and
intentionally retains the established EasySpin program names.\cite{ref1,ref2}
Therefore, \emph{garlic}, \emph{chili}, \emph{pepper}, \emph{salt},
\emph{cardamom}, \emph{saffron}, \emph{spidyan}, and \emph{curry}
identify TorchSpin ports of the corresponding EasySpin simulation
pathways. We evaluated representative calculations from each pathway
against independently generated EasySpin outputs using matched
spin-system, experimental, and numerical parameters. Each comparison
used identical field, frequency, time, or temperature axes without
interpolation. One scalar amplitude factor was applied over each
complete TorchSpin trace before calculating the normalized residual.

\begin{figure}[htbp]
\centering
\includegraphics[width=6.07576in,height=0.64\textheight,keepaspectratio]{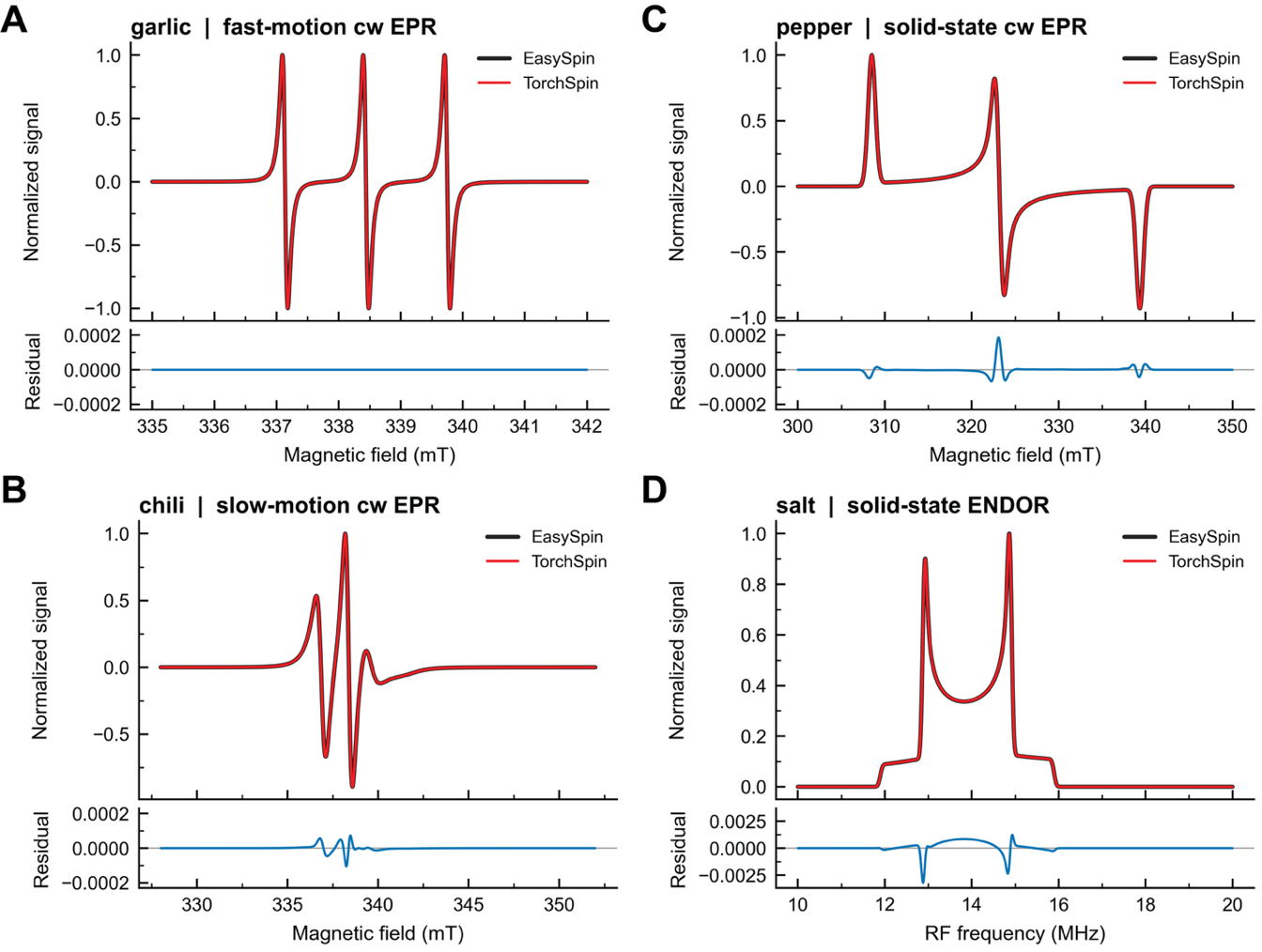}
\caption{Numerical agreement between TorchSpin and EasySpin across representative
EPR and ENDOR simulations. TorchSpin (red) and independently generated
EasySpin (black) spectra are compared for (\textbf{A}) isotropic
fast-motion cw EPR using \emph{garlic}, (\textbf{B}) slow-motion cw EPR
using \emph{chili}, (\textbf{C}) solid-state cw EPR using \emph{pepper},
and (\textbf{D}) solid-state ENDOR using \emph{salt.} Blue traces show
the residuals (TorchSpin − EasySpin). Each EasySpin spectrum was
normalized to its maximum absolute amplitude.}
\label{fig:1}
\end{figure}

Figure 1 compares four representative spectra from commonly used
spectral simulations. The fast-motion continuous wave (cw) EPR spectrum
calculated with \emph{garlic} was reproduced to numerical precision,
with a normalized root mean square error (NRMSE) below
10\textsuperscript{-15} (\textbf{Fig. 1A}). The calculation used a
simplified Fremy's salt model consisting of an electron spin coupled to
a single \textsuperscript{14}N nucleus through an isotropic hyperfine
interaction.\cite{ref13} This coupling produces the characteristic
three-line spectrum, while Lorentzian broadening and first-harmonic
detection generate the derivative lineshape shown in the figure. Because
no rotational correlation time was specified, the calculation represents
the isotropic solution limit and does not require powder-orientation
averaging. The stored reference was generated specifically for the
TorchSpin validation suite by running EasySpin's \emph{garlic} routine
with the same spin-system and acquisition parameters. The identical
field grids and near-zero residual establish agreement in resonance
positions, relative intensities, convolutional broadening, and
derivative conventions for this isotropic case.

The slow-motion \emph{chili} and solid-state \emph{pepper} calculations
gave NRMSE values of 7.23 x 10\textsuperscript{-5} and 9.09 x
10\textsuperscript{-5}, respectively (\textbf{Fig. 1B,C}). Their
residuals were concentrated around the steepest spectral features, while
the resonance positions and overall lineshapes remained closely aligned.
Slow-motion cw EPR provides an important comparison because rotational
diffusion and spin evolution must be treated simultaneously rather than
approximated by either the isotropic or rigid limit \cite{ref14,ref15}. The
synthetic \emph{chili} benchmark describes a nitroxide electron spin
coupled to a single \textsuperscript{14}N nucleus through anisotropic
\emph{g} and hyperfine tensors and undergoing isotropic Brownian
rotational diffusion with a rotational correlation time of 3 ns
(\textbf{Fig. 1B}). On this timescale, molecular reorientation only
partially averages the magnetic anisotropy, producing a lineshape
intermediate between an isotropic solution spectrum and a rigid-limit
powder pattern. Both programs treated these coupled dynamics using the
stochastic Liouville equation with the same orientational-basis
truncation. The EasySpin reference was generated specifically for the
TorchSpin validation suite using EasySpin's specialized fast Liouvillian
construction, whereas TorchSpin evaluated the same physical model
through a general Liouvillian pathway. The resulting spectra gave a
cosine similarity \textgreater{} 0.999 and an NRMSE of 7.23 x
10\textsuperscript{-5}, demonstrating agreement in the spin and
rotational superoperators, orientational-basis representation,
field-sweep calculation, convolutional broadening, and first-derivative
detection.

Whereas \emph{chili} tests the coupling between molecular motion and
spin evolution, \emph{pepper} tests the orientation dependence of the
static spin Hamiltonian. \emph{Pepper} is also one of EasySpin's
principal high-level simulation routines, covering rigid-limit cw EPR
calculations for powders, frozen solutions, glasses, and single
crystals. The benchmark was taken directly from EasySpin's
\emph{pepper\_rhombiclw} regression test and describes an electron spin
with a strongly rhombic \emph{g}-tensor and no coupled magnetic nuclei
(\textbf{Fig. 1C}). Its three principal \emph{g} values produce distinct
powder turning points across the field-swept spectrum, making the
lineshape sensitive to orientation-dependent resonance positions,
transition intensities, and numerical powder integration. Both
calculations used matrix diagonalization followed by symmetry-reduced
powder averaging, orientational interpolation, isotropic Gaussian
broadening, and first-harmonic detection. The spectra gave a cosine
similarity of 1.00 and an NRMSE of 9.09 x 10\textsuperscript{-5}. The
small residuals were concentrated near the sharp powder features, where
minor differences in resonance placement or interpolation are most
visible, while the overall lineshape and relative intensities remained
effectively identical.

The benchmark for solid-state ENDOR was adapted from EasySpin's
\emph{salt\_invariance\_axial1} regression test and describes an
electron with a rhombic \emph{g} tensor {[}2.00, 2.01, 2.02{]} coupled
to a single \textsuperscript{1}H nucleus through an axial hyperfine
tensor (\emph{A}) {[}-2, -2, 4{]} MHz (\textbf{Fig. 1D}). At a fixed
magnetic field of 326.5 mT and a microwave frequency of 9.7 GHz, the
orientation-dependent hyperfine and nuclear Zeeman interactions produce
a powder ENDOR envelope with distinct turning points around the proton
Larmor frequency. Both calculations used the same 1024-point
radiofrequency axis, 0.1 MHz Gaussian broadening, and the orientational
grid and interpolation settings stored with the EasySpin reference. The
resulting spectra gave a cosine similarity of 1.00 and an NRMSE of 1.92
x 10\textsuperscript{-3} with the residual concentrated near the sharp
powder features. A global amplitude scaling coefficient of 3.65 was
required to place the TorchSpin and EasySpin traces on the same scale.
This comparison therefore validates the ENDOR transition frequencies and
normalized powder lineshape due to differences in the absolute intensity
conventions used by the two implementations.

The trajectory-based \emph{cardamom} calculation reproduced the cw EPR
envelope with a cosine similarity greater than 0.999 and an NRMSE of
0.0201 (\textbf{Fig. S1}). Unlike the deterministic calculations in
\emph{garlic}, \emph{chili}, \emph{pepper}, and \emph{salt,} the
EasySpin and TorchSpin spectra were generated from independent finite
trajectory ensembles. The comparison used a synthetic nitroxide (S=1/2)
with anisotropic \emph{g}-values and \textsuperscript{14}N hyperfine
tensors undergoing unrestricted isotropic Brownian rotational diffusion
with a rotational correlation time of 1 ns. Each trajectory describes
the changing orientation of the nitroxide molecular frame relative to
the laboratory field, allowing the anisotropic magnetic-interaction
tensors to be updated during propagation of the spin signal. For each of
100 powder orientations, 100 rotational trajectories were propagated for
500 steps using 0.1 ns spatial-dynamics and spin-propagation time steps.
The EasySpin reference was generated specifically for this study by
running the EasySpin \emph{cardamom} routine with a fixed random seed
using nitroxide tensor values adapted from a standard EasySpin test
system through a script included in the TorchSpin validation suite.
TorchSpin generated a separate trajectory ensemble using its own
random-number generator, so the residual contains finite-ensemble
sampling variation in addition to any differences between the two
implementations.

TorchSpin's \emph{saffron} implementation reproduced the EasySpin
time-domain two-pulse ESEEM trace, including both the slower modulation
envelope and the rapid nuclear-frequency oscillations, with an NRMSE of
0.0101 (\textbf{Fig. S2}). This synthetic benchmark was taken directly
from EasySpin's \emph{saffron\_2psimple} regression test rather than
constructed specifically for TorchSpin. It describes a powder-averaged
electron spin coupled to a single \textsuperscript{14}N nucleus through
isotropic hyperfine and rhombic nuclear-quadrupole interactions and
probed using an ideal \((\pi/2 - \tau - \pi - \tau - echo)\ \)sequence
at a fixed magnetic field.\cite{ref16,ref17} Hyperfine and quadrupole evolution
within the two electron-spin manifolds generates multiple nuclear
transition frequencies, whose interference produces the slow envelope
and faster oscillations visible in the trace. The EasySpin and TorchSpin
calculations used matched spin-system, pulse-timing, acquisition, and
powder-integration settings and were compared on identical time axes
without interpolation. The mean-centered cosine similarity exceeded
0.999, confirming that agreement remained high for the modulation
pattern itself after removal of the nonzero signal baseline.

Agreement also extended beyond conventional spectral lineshapes. Whereas
\emph{saffron} validates a predefined pulse experiment, \emph{spidyan}
tests direct propagation of a user-defined shaped pulse sequence. The
benchmark was adapted from EasySpin's electron--nucleus \emph{spidyan}
regression case and used an electron with g= 2.00232 coupled to a single
\textsuperscript{1}H nucleus through an anisotropic hyperfine tensor
(8,45,45 MHz), at 1195 mT and 33.5 GHz (\textbf{Fig. S3}). A 0.1 µs
\emph{π} -pulse with a quarter-sine amplitude profile, a 0.015 µs rise
time, and a linear frequency sweep from -100 to 100 MHz was followed by
0.5 µs of free evolution. The electron \emph{z}-magnetization was
recorded throughout both events, providing a direct test of waveform
generation, coupled electron--nuclear spin propagation, numerical time
integration, and operator detection. The complete 6002-point
trajectories gave a cosine similarity of 1.00 and an NRMSE of 6.39 x
10\textsuperscript{-5}, with a maximum normalized residual of 3.48 x
10\textsuperscript{-4}. The temperature-dependent effective magnetic
moment of an exchange-coupled Gd(III)--Cu(II) system calculated with
\emph{curry} gave an NRMSE of 1.48 x 10\textsuperscript{-7} after
converting the TorchSpin molar susceptibility to µ\textsubscript{eff}
(\textbf{Fig. S4}). Together, these comparisons establish numerical
correspondence across the eight tested TorchSpin ports and several
classes of magnetic-resonance observables.

\begin{figure}[htbp]
\centering
\includegraphics[width=6.40055in,height=0.64\textheight,keepaspectratio]{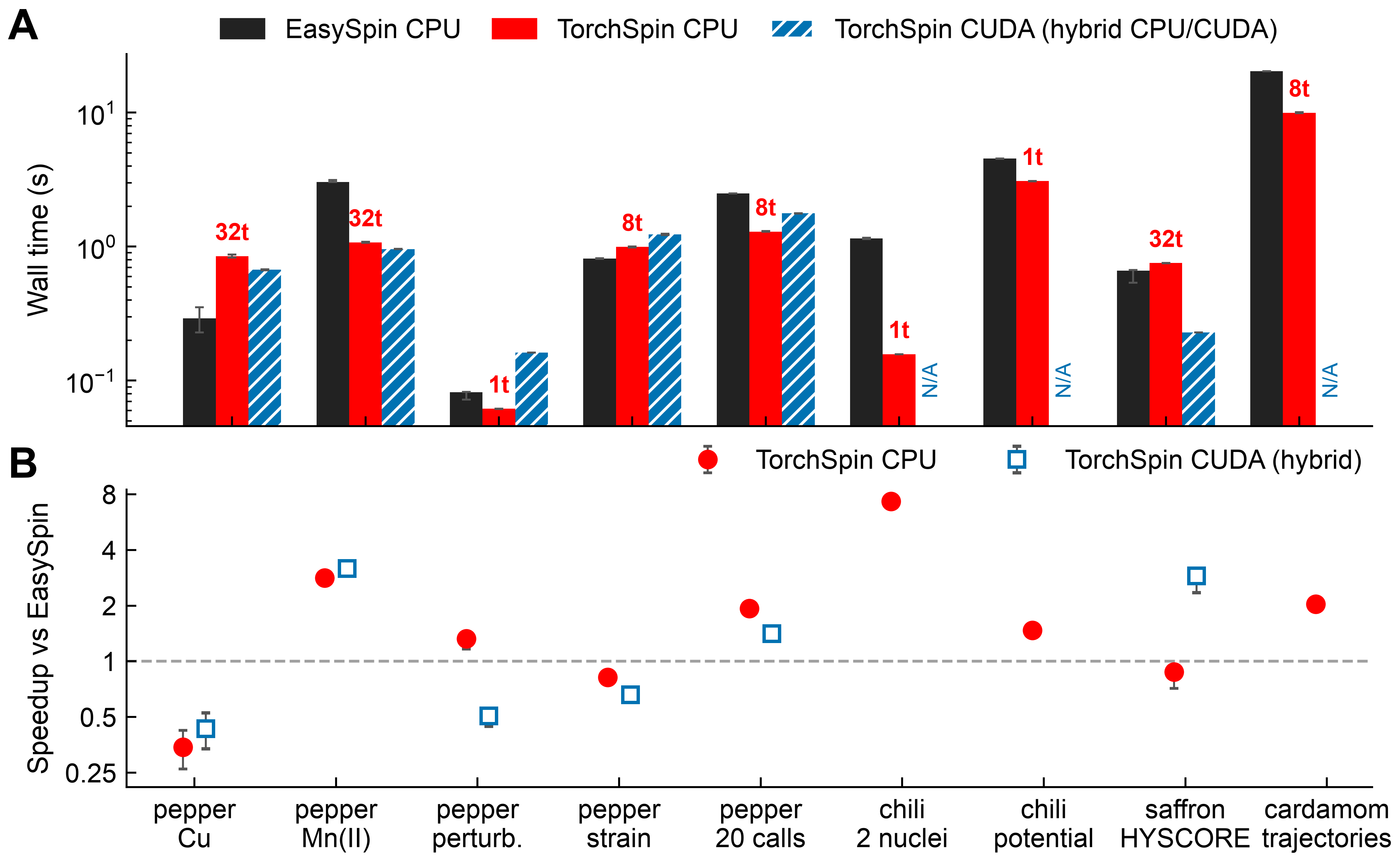}
\caption{Performance of representative EasySpin and TorchSpin simulation
workloads on a common computing platform. (\textbf{A}) Median wall time
for EasySpin on the CPU, TorchSpin using the selected CPU thread count,
and TorchSpin in CUDA mode. TorchSpin CPU calculations were evaluated
using 1, 8, and 32 threads; red labels identify the selected
configuration, defined as the smallest thread count whose timing
interquartile range (IQR) overlapped that of the configuration with the
lowest median time. Hatched blue bars denote hybrid CUDA execution, in
which verified GPU kernels are combined with CPU-side operations. N/A
indicates that the corresponding \emph{chili} or \emph{cardamom}
implementation does not currently provide a CUDA execution path.
(\textbf{B}) Speedup relative to EasySpin, calculated as the EasySpin
wall time divided by the corresponding TorchSpin wall time. Values above
unity indicate faster TorchSpin execution, and the dashed line marks
equal performance. Points show ratios of median times; error bars are
conservative IQR-propagated bounds calculated from the EasySpin and
TorchSpin measurements. Each condition comprised one untimed warm-up
followed by five timed replicates on a local workstation; bars and
points report medians, and error bars span the IQR.}
\label{fig:2}
\end{figure}

\subsection*{Performance is workload- and hardware-dependent}

Performance was evaluated using nine matched EasySpin and TorchSpin
workloads chosen to represent distinct computational demands rather than
a single favorable simulation regime. These included matrix and
perturbative \emph{pepper} calculations, strain broadening, repeated
forward calculations representative of fitting, two stochastic Liouville
\emph{chili} calculations, two-dimensional HYSCORE with \emph{saffron},
and trajectory-based simulation with \emph{cardamom}. All calculations
were performed sequentially on the same workstation using a common
32-core CPU allocation. TorchSpin was additionally evaluated on an
NVIDIA RTX 4090 for the six workloads with CUDA-capable implementations.
Each condition included one untimed warm-up followed by five timed
replicates. Complete spin systems, numerical settings, and benchmark
procedures are provided in the Supplementary Information.

TorchSpin CPU execution was faster than EasySpin for six of the nine
workloads (\textbf{Fig. 2}). The largest difference was observed for the
two-nucleus slow-motion \emph{chili} calculation, for which TorchSpin
was 7.32-fold faster. TorchSpin was also faster for the Mn(II) matrix
calculation (2.83-fold), trajectory-based \emph{cardamom} calculation
(2.04-fold), 20-call \emph{pepper} fitting pattern (1.93-fold),
orienting-potential \emph{chili} calculation (1.47-fold), and dense
perturbative \emph{pepper} calculation (1.33-fold). For the short
perturbative calculation, the IQR-derived bounds were 1.16--1.35-fold,
reflecting the greater relative influence of fixed overhead and timing
variation on simulations requiring approximately 0.06 - 0.08 s. EasySpin
remained faster for the Cu(II) matrix powder, strain-summation, and CPU
HYSCORE calculations, for which the EasySpin/TorchSpin time ratios were
0.344, 0.820, and 0.877, respectively. TorchSpin was therefore
competitive across these benchmarks but did not uniformly outperform
EasySpin.

CPU scaling depended strongly on the structure of the calculation
(\textbf{Fig. S5} and \textbf{Table S1}). Increasing the thread count
substantially accelerated the matrix-based Cu(II) and Mn(II) powders,
strain summation, HYSCORE, and trajectory propagation. Additional
threads provided little benefit for the \emph{chili} calculations and
increased the wall time of the short perturbative workload. The 20-call
fitting pattern improved from one to eight threads but became markedly
slower at 32 threads, showing that thread-management overhead can
outweigh the benefit of additional parallelism. For each workload,
Figure 2 therefore reports the smallest tested thread count whose timing
IQR overlapped that of the configuration with the lowest median wall
time.

CUDA execution was verified for the five \emph{pepper} workloads and the
\emph{saffron} HYSCORE calculation. GPU kernels were observed in all six
cases, but they accounted for no more than 16.6\% of the profiled wall
time, and substantial portions of each calculation remained on the CPU.
In particular, eigen-decompositions of the Cu(II) and Mn(II) powders
were routed through the CPU pool. These measurements are therefore
designated hybrid CPU/CUDA execution. CUDA reduced the TorchSpin wall
time for the Cu(II), Mn(II), and HYSCORE workloads but was slower than
the selected TorchSpin CPU configuration for the perturbative, strain,
and repeated-call calculations. The largest GPU benefit occurred for
HYSCORE, which decreased from 0.752 s on the selected CPU configuration
to 0.228 s in CUDA mode and was 2.89-fold faster than EasySpin. By
contrast, CUDA increased the time required for the perturbative
calculation from 0.0616 to 0.162 s, consistent with device-transfer and
kernel-launch overhead dominating a short simulation. The current
\emph{chili} and \emph{cardamom} implementations use CPU-resident SciPy
or NumPy operations and consequently have no CUDA result yet.

Neither additional CPU threads nor CUDA execution therefore provides a
universal advantage for an individual EPR simulation. Acceleration
depends on whether the dominant numerical operations can be parallelized
efficiently enough to offset thread-management, data-transfer, and
kernel-launch costs. The practical value of the tensor implementation
nevertheless extends beyond forward-simulation speed. Supported
simulations can participate directly in differentiable optimization and
machine-learning workflows, where batched evaluation and backward
propagation provide a more consequential use of accelerator hardware.

\subsection*{Differentiable spectral feedback improves learned EPR inversion}

Forward-simulation speed alone does not capture the main consequence of
implementing EPR calculations in a differentiable tensor framework.
Conventional fitting treats each spectrum as a new optimization problem.
In contrast, a neural inverse model can learn a reusable mapping from
spectra to spin Hamiltonian parameters across a defined family of
systems, after which each new spectrum can be analyzed by a single
forward pass.\cite{ref18} TorchSpin then provides a physics-based test of
that prediction: the inferred parameters are returned to the
differentiable simulator, and the disagreement between the reconstructed
and target spectra can refine either the prediction or the network
itself (\textbf{Fig. 3}) \cite{ref9,ref10}. We therefore evaluated TorchSpin in
a proof-of-concept workflow that separates simulation-trained parameter
estimation from simulator-in-the-loop spectral refinement.

To test whether TorchSpin could connect an EPR forward model directly to
a learned inverse model, we trained a simple one-dimensional
convolutional neural network (CNN) to predict the principal values of
the \emph{g} and \textsuperscript{63}Cu hyperfine tensors and a Gaussian
linewidth from noisy Cu(II) powder spectra (\textbf{Fig. 3A} and
\textbf{Table S2}). This example is meant to serve as an illustration of
the opportunities provided by this platform rather than a comprehensive
science case. The independently generated EasySpin library contained
4,096 training, 512 validation, and 512 test spectra spanning axial and
rhombic, blue-copper-like parameter ranges\cite{ref19,ref20,ref21}. Each spectrum
contained Gaussian noise with a standard deviation drawn uniformly from
1--10\% of the maximum absolute signal. The CNN was first trained by
supervised regression against the seven generating parameters. We then
refined the network through TorchSpin using a spectral objective that
combined derivative-spectrum agreement, integrated-spectrum agreement,
and peak-weighted reconstruction error (\textbf{Fig. 3A}). Parameter
labels were excluded from both the refinement objective and refinement
checkpoint selection, allowing us to assess the effect of spectral
feedback after supervised initialization. This comparison therefore
tested whether spectral feedback from a differentiable simulator can
improve an already trained inverse model without an explicit
parameter-error penalty.

\begin{figure}[htbp]
\centering
\includegraphics[width=4.11421in,height=0.64\textheight,keepaspectratio]{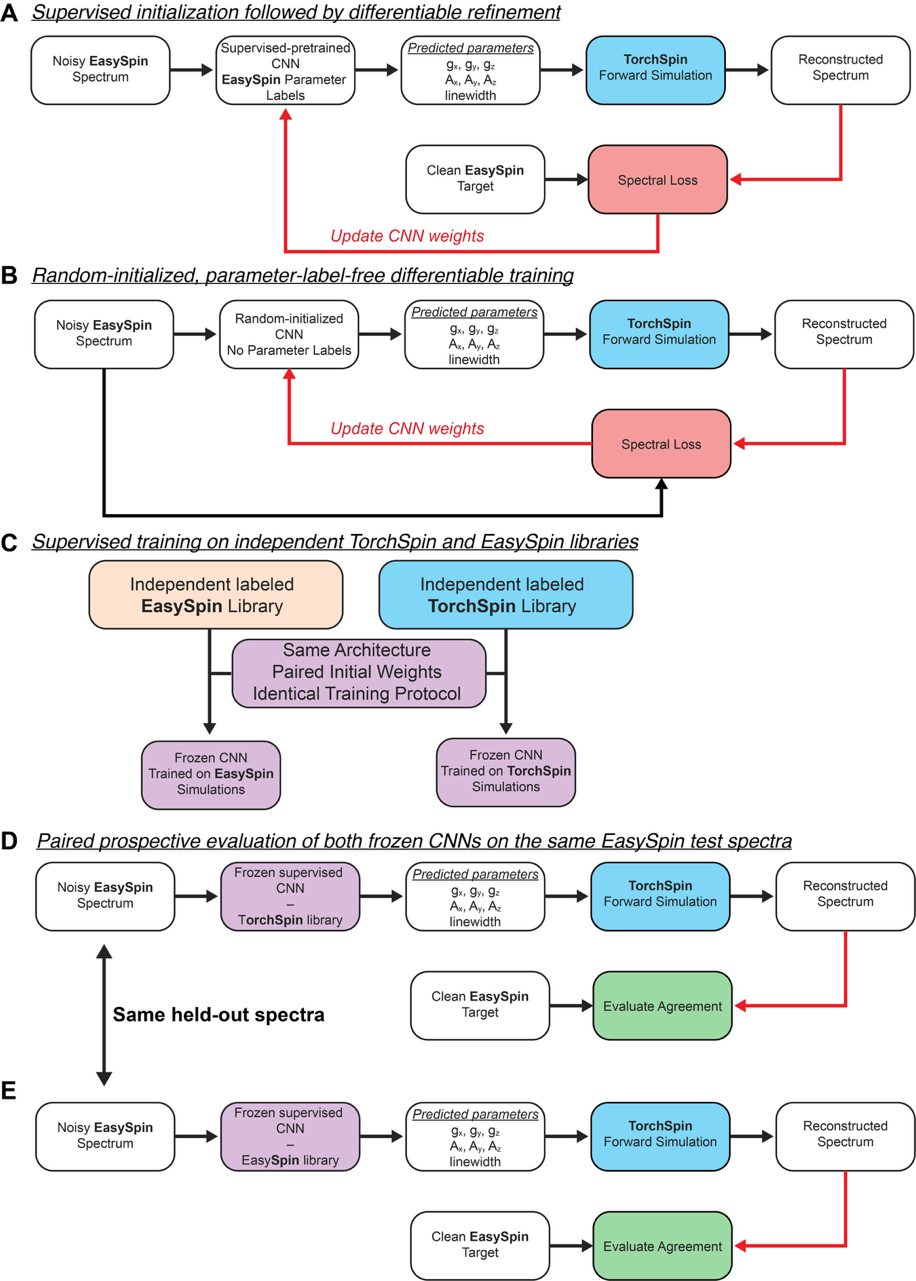}
\caption{Overview of the CNN training and evaluation workflows used to test
TorchSpin. (\textbf{A}) A CNN was first trained by supervised regression
against the seven generating parameters of an EasySpin spectral library.
During subsequent differentiable refinement, the CNN mapped noisy
EasySpin spectra to the principal values of the \emph{g} and
\textsuperscript{63}Cu hyperfine tensors and a Gaussian linewidth.
TorchSpin reconstructed the predicted spectrum, and a spectral loss
against the corresponding clean EasySpin target was backpropagated
through the simulator to update the CNN. (\textbf{B}) In the
parameter-label-free experiment, independently initialized CNNs were
trained without supervised pretraining, parameter labels, or clean
spectral targets. The same fixed noisy EasySpin spectrum served as both
the CNN input and the reconstruction target, and TorchSpin supplied the
spectral gradients used to update the network. (\textbf{C}) For the
cross-engine library comparison, supervised CNNs were trained on
independent labeled libraries generated with EasySpin or TorchSpin. The
two conditions used the same architecture and training protocol, with
identical initial weights within each paired initialization seed. The
resulting checkpoints were frozen before prospective evaluation.
(\textbf{D,E}) CNNs trained on the TorchSpin (\textbf{D}) or EasySpin
(\textbf{E}) library were evaluated on the same held-out noisy EasySpin
spectra. Predictions from both conditions were reconstructed using the
same TorchSpin forward model and compared with the corresponding clean
EasySpin spectra. Black arrows denote input and forward-computation
paths; red arrows denote spectral-error propagation or comparison.}
\label{fig:3}
\end{figure}

Differentiable spectral refinement improved 373 of the 512 test
reconstructions (72.9\%). Median cosine similarity to the corresponding
clean EasySpin spectrum increased from 0.977 before refinement to 0.985
afterward (\textbf{Fig. 4A,B} and \textbf{Table S3}), with the largest
gains among initially poorer reconstructions. The 10th-percentile
similarity increased from 0.914 to 0.951, the minimum increased from
0.593 to 0.818, and the fraction reaching a similarity of at least 0.950
increased from 76.6\% to 90.2\%. Agreement also improved within each
predefined noise interval (\textbf{Fig. S6}).

Improved spectral reconstruction was not accompanied by uniform
improvement in the recovery of the generating parameters. Refinement
reduced the prediction errors for \emph{g\textsubscript{x}} and
\emph{g\textsubscript{y}}, whereas errors in \emph{g\textsubscript{z}}
and the \textsuperscript{63}Cu hyperfine components increased modestly.
The clearest tradeoff occurred for Gaussian linewidth, for which the
mean absolute error increased from 0.0736 to 0.0921 mT after refinement,
which seems small but represents an increase of approximately 25\%.
Thus, the spectrum-level objective identified parameter combinations
that reproduced the target more closely without necessarily converging
toward the original generating vector. This compensation illustrates
that differentiable spectral feedback improves the physical consistency
of the reconstructed spectrum but does not, by itself, resolve parameter
correlations or guarantee a unique spin Hamiltonian. Once trained, the
CNN predicted all seven parameters from a single GPU-resident spectrum
in a median of about 1 ms, while generation of the corresponding
512-point TorchSpin spectrum required approximately 70 ms.

We next asked whether differentiable spectral feedback could train an
inverse model without parameter supervision. Three CNNs were initialized
independently from random weights and trained using a newly generated
EasySpin library containing 4,096 training and 512 validation spectra
(\textbf{Fig. 3B}). The generating spin-Hamiltonian parameters and clean
spectra were withheld throughout optimization and checkpoint selection.
For each example, the CNN predicted a bounded seven-parameter
representation, TorchSpin reconstructed the corresponding spectrum, and
the resulting spectral error was backpropagated through the simulator to
update the network. The same fixed noisy spectrum therefore served as
both the CNN input and reconstruction target. Unlike the preceding
refinement experiment, these networks received neither supervised
initialization nor a parameter-based loss, while a separately trained
supervised CNN provided a parameter-regression control.

Parameter-label-free training produced similar central-distribution
performance across the three independently initialized models. Median
cosine similarity to the clean EasySpin test spectra was 0.992, 0.992,
and 0.991 for the three models, compared with 0.970 for the supervised
control evaluated on the same prospective test set (\textbf{Fig. 4C} and
\textbf{Table S5}). Each label-free model reconstructed 98.6\% of the
test spectra with a cosine similarity of at least 0.950, whereas the
supervised control reached this threshold for 65.2\%. Spectral agreement
improved relative to the control for 88.3--91.2\% of the paired test
spectra. The central distributions were therefore reproducible across
initialization seeds, although two models retained an isolated severe
failure for the same test spectrum (\textbf{Fig. S9}). These results
show that TorchSpin supplied sufficient spectral gradients to learn
accurate reconstruction across most of the defined Cu(II) domain without
parameter labels or clean-spectrum targets (\textbf{Fig. S7)}.

\begin{figure}[htbp]
\centering
\includegraphics[width=6.50000in,height=0.64\textheight,keepaspectratio]{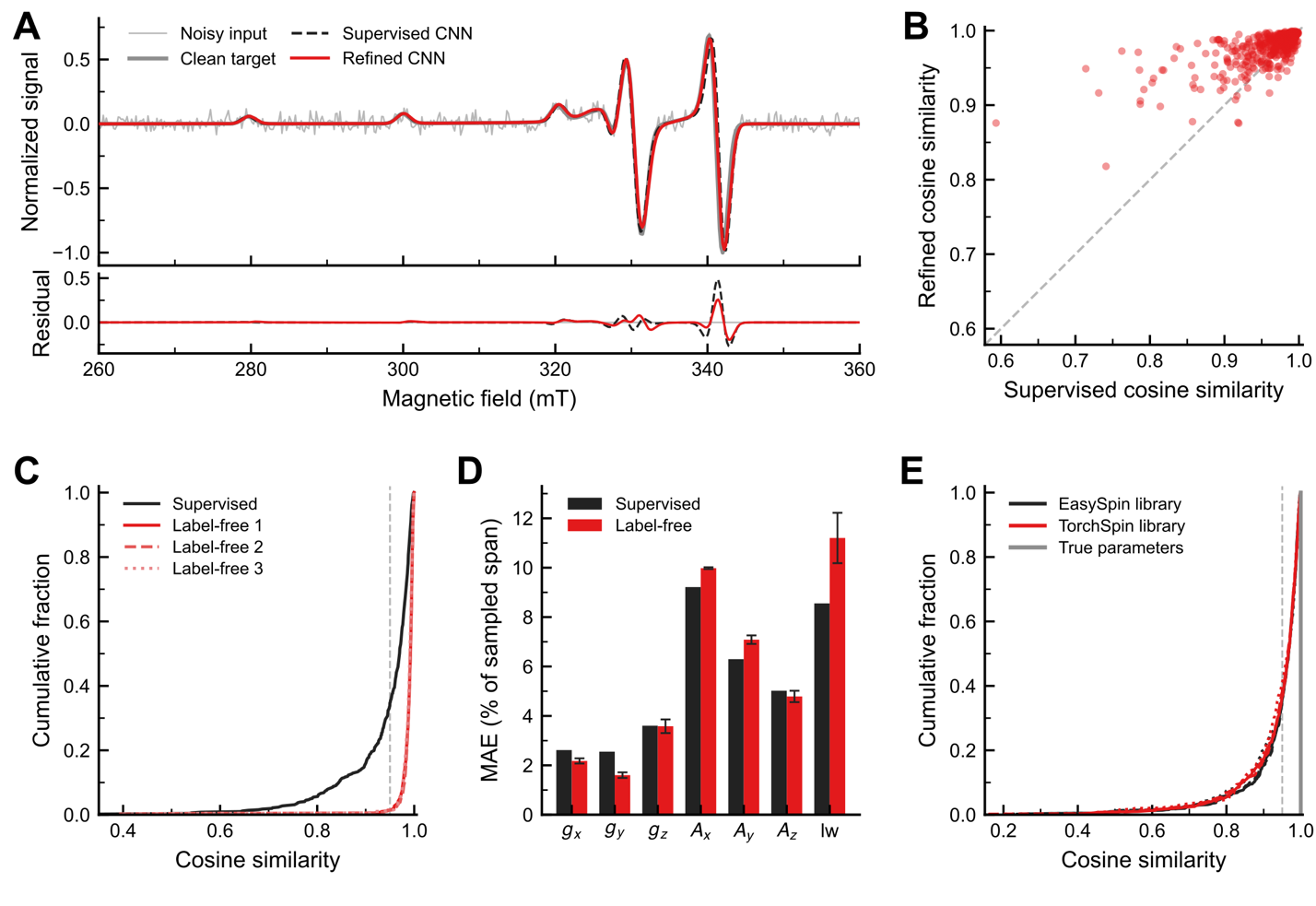}
\caption{Differentiable spectral feedback and simulation-trained CNN inversion of
Cu(II) powder EPR spectra. (\textbf{A}) Representative held-out spectrum
showing the noisy input, clean EasySpin target, and TorchSpin
reconstructions obtained from the supervised CNN before and after
differentiable spectral refinement. Residuals are shown relative to the
clean target. (\textbf{B}) Paired comparison across all 512 test
spectra. Refinement improved 373 spectra (72.9\%), increasing the median
cosine similarity from 0.977 to 0.985. The dashed diagonal denotes equal
performance before and after refinement. (\textbf{C}) Empirical
cumulative distributions of spectral agreement for the supervised
control and three independently initialized, parameter-label-free CNNs
trained from random initialization using TorchSpin spectral feedback.
All models were evaluated on the same 512-spectrum EasySpin test
library. The fraction exceeding the 0.950 threshold marked by the dashed
line was 65.2\% for the supervised control and 98.6\% for each
label-free model. (\textbf{D}) Mean absolute error (MAE) in the
predicted \emph{g\textsubscript{x}}, \emph{g\textsubscript{y}},
\emph{g\textsubscript{z}}, \emph{A\textsubscript{x}},
\emph{A\textsubscript{y}}, \emph{A\textsubscript{z}}, and Gaussian
linewidth (\emph{lw}), expressed as a percentage of the full
prespecified sampling span for each parameter. Label-free bars report
the mean across three independently initialized models, and error bars
indicate the sample standard deviation. (\textbf{E}) Cross-engine
comparison of supervised CNNs trained on independent labeled EasySpin or
TorchSpin spectral libraries using paired initial weights and identical
training protocols. The frozen models were evaluated on the same
held-out EasySpin spectra, and all predicted parameters were
reconstructed using TorchSpin. Color identifies the training-library
simulator, line style distinguishes the three paired initialization
seeds, and the gray curve shows TorchSpin reconstructions generated from
the true sampled parameters. The dashed vertical line marks a cosine
similarity of 0.950. Spectral agreement in all panels was evaluated
against the corresponding noise-free EasySpin spectrum.}
\label{fig:4}
\end{figure}

The increase in spectral agreement did not correspond to uniformly
improved recovery of the generating Hamiltonian (\textbf{Fig. 4D}).
Relative to the supervised control, all three label-free models slightly
reduced the mean absolute error for \emph{g\textsubscript{x}},
\emph{g\textsubscript{y}}, and \emph{A\textsubscript{z}}, but slightly
increased the error for \emph{A\textsubscript{x}},
\emph{A\textsubscript{y}} and linewidth; recovery of
\emph{g\textsubscript{z}} depended on initialization. The models
predicted varying values for every parameter, and fewer than 1\% of
predictions fell within 2\% of the allowed parameter limits, which
suggests that the networks had not simply learned to return fixed values
or values at those limits. Nevertheless, 110 of 1,536 same-input
cross-seed comparisons combined a reconstruction cosine of at least
0.995 with a difference of at least 10\% of one parameter span.
Differentiable spectral feedback therefore constrained the networks to
physically productive regions of the bounded parameter space without
establishing a unique Hamiltonian representation (\textbf{Fig. S8,S9}
and \textbf{Table S5}).

A complementary use of TorchSpin is to generate labeled spectral
libraries for conventional supervised learning. We therefore asked
whether a CNN trained exclusively on TorchSpin simulations could analyze
spectra generated independently by EasySpin without retraining or
adaptation (\textbf{Fig. 3C, D and E}). Three CNNs were trained on
independently sampled TorchSpin libraries, and three matched controls
were trained on EasySpin libraries using identical architectures,
optimization schedules, and paired initial weights. Training and
validation contained 4,096 and 512 spectra, respectively, whereas
evaluation used a prospectively isolated set of 512 noisy EasySpin
spectra. The parameter combinations were sampled independently across
libraries, and no exact parameter or spectrum duplicates were detected
between the test set and either training source. Predictions from both
conditions were reconstructed with TorchSpin during evaluation, allowing
the source of the supervised training library to be compared against a
common physical decoder.

The source of the training library had little effect on the central
performance of the inverse model. Across the three paired initialization
seeds, the median cosine similarity to the clean EasySpin test spectra
averaged 0.966 for the TorchSpin-trained models and 0.967 for the
EasySpin-trained references (\textbf{Fig. 4E}). The corresponding
fractions of spectra reaching a cosine similarity of at least 0.950 were
61.6\% and 63.9\%, respectively. Transfer was modestly poorer in the
lower tail, where the mean lower-decile similarity was 0.843 after
TorchSpin training and 0.865 after EasySpin training, and both
conditions retained severe individual failures. Nevertheless, the paired
mean difference in cosine similarity was only −0.00213 to −0.00339
across the three seeds, and the component-wise parameter errors differed
by no more than 0.49 percentage points of the sampled parameter span. At the exact test parameters, TorchSpin reproduced
the corresponding EasySpin spectra with a median NRMSE of
1.24x10\textsuperscript{-5}, indicating that the remaining
reconstruction error arose primarily from parameter inference rather
than disagreement between the forward calculations (\textbf{Fig.
S10-S13} and \textbf{Tables S6, S7}).

\FloatBarrier
\section*{Discussion}

The central contribution of TorchSpin is a Python/PyTorch-based
implementation that makes established EPR calculations available within
differentiable, tensor-based workflows \cite{ref12}. Across representative
continuous-wave EPR, ENDOR, pulse EPR, spin-dynamics, and magnetometry
calculations, TorchSpin reproduced the corresponding EasySpin reference
spectra while retaining the familiar organization and nomenclature of
the EasySpin framework \cite{ref1}. Its forward-simulation performance was
competitive but depended on the structure of the calculation, with
neither additional CPU threads nor CUDA execution providing a universal
advantage. The toy-example blue-copper learning experiments demonstrate
why the tensor implementation remains useful beyond raw simulation
speed: the forward calculation can serve as a differentiable physical
constraint, a decoder for parameter-label-free learning, or a generator
of synthetic libraries for reusable inverse models.

The distinction between parameter supervision and spectral supervision
is central to interpreting these results. In supervised training, the
model learns to predict the known parameters used to generate each
spectrum. In spectral training, it learns to predict parameters that
reproduce the spectrum, even if they differ from the original values.
These objectives converge when the inverse problem is well constrained,
but they can diverge when several parameter combinations produce similar
spectra. The parameter-label-free models reconstructed spectra
accurately but did not consistently recover every generating parameter.
Differences between independently trained models that produced similar
spectra were concentrated primarily in linewidth, suggesting that
broadening was less reliably constrained under the conditions tested.
Differentiable simulation therefore does not eliminate the need for
physically meaningful parameter bounds, independent experimental
constraints, or uncertainty analysis. Instead, it provides a way to
expose how strongly the measured spectrum constrains each parameter,
enables mechanisms for parameter uncertainty estimation, and how to
incorporate additional information into the same optimization framework.

These learned inverse models are most useful as complements to
conventional fitting rather than replacements for it. Once trained, the
CNN converts a spectrum into an initial set of spin-system parameters in
a single forward pass, avoiding a new iterative search for every
measurement. On our hardware, the final CNN required a median of 0.976
ms to analyze one spectrum using a single CPU thread and 1.05 ms on an
RTX 4090 when spectrum normalization and host-to-device transfer were
included. GPU-resident inference required 1.02 ms for one spectrum,
whereas a batch of 512 spectra was processed in 1.27 ms. Generating the
corresponding 512-point TorchSpin reconstruction required an additional
68.4 ms (\textbf{Table S8}). The network can therefore rapidly screen
large spectral series or provide informed starting parameters, after
which TorchSpin or a conventional fitting procedure can be used
selectively to test spectral agreement, refine the parameters, and
assess uncertainty.

The cross-engine experiment illustrates a second, complementary role for
TorchSpin. A simulator used only to generate labeled training spectra
does not need to be differentiable, but a tensor-based implementation
allows spectral libraries to be generated and consumed within the same
Python environment used for model training. CNNs trained on TorchSpin
spectra transferred without adaptation to independently generated
EasySpin spectra with median reconstruction accuracy similar to that of
the EasySpin-trained controls. Under the tested Cu(II) conditions, the
learned inverse mapping was therefore not strongly dependent on
numerical features unique to the TorchSpin simulations. The more
demanding test will be whether models trained with TorchSpin remain
accurate in the presence of experimental features like baseline
distortions, phase errors, field-calibration uncertainty, unresolved
mixtures, and broadening mechanisms that were not represented in the
present libraries.

\begin{figure}[htbp]
\centering
\includegraphics[width=6.50000in,height=0.64\textheight,keepaspectratio]{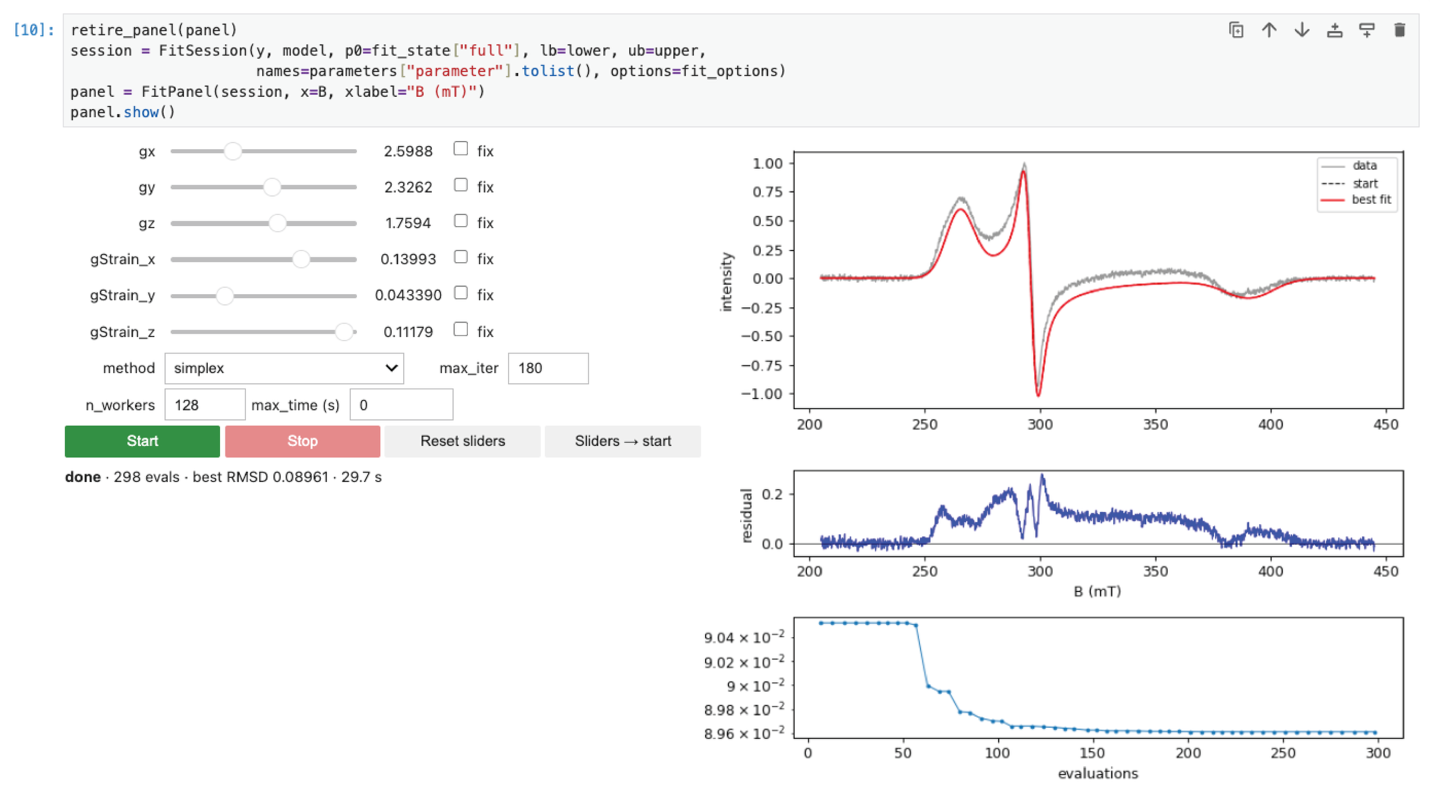}
\caption{Interactive Jupyter interface for conventional cw EPR fitting with
TorchSpin. Parameter sliders and fixed-parameter controls allow
adjustment of the spin system parameters. Additional controls select the
optimization method and fitting limits. The plots display the input
spectrum, starting simulation, best fit, residual, and best RMSD as a
function of model evaluations. The interface supports repeated
adjustment and refitting within the notebook.}
\label{fig:5}
\end{figure}

TorchSpin also supports conventional cw EPR fitting through an
interactive Jupyter interface modeled on the workflow of EasySpin's
\emph{esfit} (\textbf{Fig. 5}). Users can adjust initial parameter
values, fix selected parameters, choose an optimization method, and
monitor the fitted spectrum, residuals, and fitting progress within the
notebook. The interface connects these controls to TorchSpin forward
simulations, allowing supported calculations to use CPU or hybrid
CPU/CUDA execution according to the selected model and hardware. As the
benchmarks demonstrate, the resulting performance benefit depends on the
workload.

Within these limits, TorchSpin provides a practical route for bringing
established EPR simulation into the PyTorch computational environment.
Its organization and program names are derived directly from EasySpin,
preserving a familiar framework while exposing supported calculations to
tensor batching, automatic differentiation, and machine-learning tools.
The principal advantage is therefore not a uniform reduction in the time
required for every forward calculation. Rather, the same physical model
can generate training data, evaluate predicted parameters, and propagate
spectral disagreement through an optimization or learning system. The
accompanying code, validation data, benchmarks, and notebooks provide a
reproducible foundation for extending this approach to additional EPR
methods and, importantly, for testing it against experimental spectra.

\FloatBarrier
\section*{Conclusions}

TorchSpin brings representative EPR simulation methods into a
differentiable PyTorch framework while preserving the organization and
nomenclature familiar to EasySpin users. The validation calculations
establish close numerical correspondence across continuous-wave EPR,
ENDOR, pulse EPR, spin dynamics, and magnetometry, whereas the
benchmarks define the present computational tradeoffs: TorchSpin is
competitive across diverse workloads, but neither CPU parallelism nor
hybrid CUDA execution provides a universal speed advantage. The more
consequential capability is that supported forward calculations can
participate directly in automatic differentiation. In the Cu(II)
demonstration, this allowed spectral disagreement to refine a supervised
inverse model and to train models from random initialization without
parameter labels or clean spectral targets. Separately, CNNs trained
using TorchSpin-generated libraries transferred to independently
generated EasySpin spectra within the tested domain, supporting the use
of TorchSpin as both a physical decoder and a synthetic-data generator.
These proof-of-concept results provide a reproducible foundation for
combining established EPR physics with rapid learned inference,
physics-constrained optimization, and future machine-learning workflows.

\FloatBarrier
\section*{Methods}

\subsection*{Software implementation}

TorchSpin was implemented in Python using PyTorch for tensor operations,
with NumPy and SciPy retained for numerical procedures that have not yet
been converted to tensor-native form. Spin systems, experimental
settings, and numerical options are represented by the SpinSystem,
Experiment, and Options objects, respectively. Magnetic-field values are
specified in mT, spin-Hamiltonian energies and interaction parameters in
MHz, and angular quantities in radians. The simulation routines and
their names, including \emph{garlic}, \emph{chili}, \emph{pepper},
\emph{salt}, \emph{saffron}, \emph{spidyan}, \emph{cardamom}, and
\emph{curry}, were ported directly from the corresponding EasySpin
framework to preserve a familiar organization for EPR users.
Differentiable calculation paths retain parameter-dependent operations
on the PyTorch computation graph, allowing gradients of the simulated
spectrum with respect to supported spin-system and experimental
variables to be obtained through automatic differentiation.

\subsection*{EasySpin reference comparisons}

Numerical correspondence was evaluated against stored EasySpin/MATLAB
reference calculations for eight representative cases spanning
continuous-wave EPR, ENDOR, pulse EPR, spin dynamics, and magnetometry.
Each TorchSpin calculation used the same spin system, experimental
coordinates, and internal numerical settings as the corresponding
reference, including the stored powder-grid or basis-truncation
parameters. Output axes were required to have the same length and agree
within floating-point tolerance; none of the comparisons used
interpolation or index-based alignment across different coordinate
grids. Before calculating the residual, a single coefficient, c =
\(\frac{\mathbf{y}_{ES} \cdot \mathbf{y}_{TS}}{\mathbf{y}_{TS} \cdot \mathbf{y}_{TS}}\),
was applied to the complete TorchSpin trace to minimize the squared
amplitude difference. Both traces were then divided by the maximum
absolute EasySpin signal. No smoothing, baseline correction, shifting,
cropping, or point exclusion was applied. Agreement was quantified using
cosine similarity, mean-centered cosine similarity, the scaled
normalized root-mean-square error,
\(NRMSE = \frac{\left\| c\,\mathbf{y}_{TS} - \mathbf{y}_{ES} \right\|_{2}}{\left\| \mathbf{y}_{ES} \right\|_{2}},\)
and the maximum absolute normalized residual.

\subsection*{Performance benchmarking}

Performance measurements were collected on a local workstation equipped
with an AMD Ryzen Threadripper PRO 5995WX processor, 503 GB of memory,
and two NVIDIA GeForce RTX 4090 GPUs. All EasySpin, TorchSpin CPU, and
TorchSpin CUDA calculations were run sequentially with process affinity
restricted to the same 32 physical CPU cores. TorchSpin CPU performance
was measured using 1, 8, and 32 threads, with the PyTorch, OpenMP, MKL,
OpenBLAS, and NumExpr thread pools set to the same value before import.
EasySpin calculations used MATLAB R2026a with a maximum of 32
computational threads. The comparison used TorchSpin v0.2.2 and an
EasySpin development version. Each workload included one untimed warm-up
followed by five timed replicates, and wall times are reported as
medians with interquartile ranges. The selected TorchSpin CPU result was
the smallest tested thread count whose timing interquartile range
overlapped that of the configuration with the lowest median wall time.
CUDA calls were synchronized immediately before and after timing, and
profiler and device audits were used to distinguish hybrid CPU/CUDA
execution from unsupported or fully GPU-resident calculations.

The nine workloads were chosen to isolate distinct computational demands
rather than to produce a single aggregate measure of performance. The
\emph{pepper} benchmarks included Cu(II) and Mn(II) matrix calculations,
a dense perturbative calculation, explicit strain summation, and a
sequence of 20 repeated forward calculations representative of iterative
fitting. The two \emph{chili} cases tested slow-motion simulations
containing either an additional coupled proton or an orienting
potential, whereas the \emph{saffron} and \emph{cardamom} benchmarks
tested two-dimensional HYSCORE and finite-trajectory propagation,
respectively. CUDA was enabled for the five \emph{pepper} workloads and
the \emph{saffron} calculation; however, these calculations were
classified as hybrid because substantial numerical work remained on the
CPU and GPU kernels accounted for no more than 17 \% of the measured
wall time. The present \emph{chili} and \emph{cardamom} implementations
rely on CPU-resident operations and were therefore reported as
unavailable in the CUDA comparison. Complete spin system descriptions
and experimental settings are provided in the Supplementary Information.

\subsection*{Cu(II) spectral libraries}

The primary Cu(II) library was generated independently with EasySpin
\emph{pepper} and contained 4,096 training, 512 validation, and 512
prospective test spectra. Each first-derivative powder spectrum
represented an (S=1/2) electron coupled to one \textsuperscript{63}Cu
nucleus and contained 512 field points spanning 260--360 mT at 9.5 GHz.
The seven target variables were the principal values
\emph{g\textsubscript{x}}, \emph{g\textsubscript{y}}, and
\emph{g\textsubscript{z}}, the corresponding hyperfine couplings
\emph{A\textsubscript{x}}, \emph{A\textsubscript{y}}, and
\emph{A\textsubscript{z}}, and a Gaussian linewidth. Parameter
combinations were sampled independently within the training, validation,
and test splits using stratified designs spanning axial and rhombic,
blue-copper-like systems. Gaussian noise was added to each spectrum with
a standard deviation sampled uniformly from 1--10\% of the maximum
absolute value of the corresponding noise-free spectrum. Both noisy and
clean spectra were retained, but only the noisy spectra were supplied as
network inputs. Model selection used the validation split, and the test
archive was not accessed for evaluation until all checkpoints had been
selected.

\subsection*{Neural inverse model and supervised training}

The inverse model was a one-dimensional residual CNN containing
1,297,127 trainable parameters. A 32-channel stem convolution was
followed by four residual downsampling blocks containing 48, 64, 96, and
128 channels, with group normalization and Sigmoid Linear Unit (SiLU)
activation functions throughout. The resulting 128 x 32 feature map was
passed through fully connected layers containing 256 and 128 hidden
units and seven outputs. A sigmoid constrained each output to the unit
interval, which was then mapped onto the sampled ranges for the
transverse \emph{g} center, \emph{g} rhombicity,
\emph{g\textsubscript{z}}, transverse hyperfine center, hyperfine
rhombicity, A\textsubscript{z}, and Gaussian linewidth (\textbf{Table
S2}). The transverse centers and rhombicities were converted to
\emph{g\textsubscript{x}}, \emph{g\textsubscript{y}},
\emph{A\textsubscript{x}}, and \emph{A\textsubscript{y }}before
simulation. Supervised training minimized the mean-squared error of the
seven normalized generating variables using AdamW, a batch size of 256,
an initial learning rate of 1.5 x 10\textsuperscript{-3}, a weight decay
of 1 x 10\textsuperscript{-5}, and cosine learning-rate decay. A new
1--10\% Gaussian-noise realization was generated from each clean
training spectrum at every epoch. Training continued for a maximum of
160 epochs with an early-stopping patience of 35 epochs, and the
checkpoint with the lowest validation parameter error was retained.

\subsection*{Differentiable spectral refinement}

The selected supervised checkpoint was refined by passing its predicted
parameters through differentiable TorchSpin \emph{pepper} and
propagating the resulting spectral error through both the simulator and
CNN. Refinement used a fixed subset of 1,024 training spectra for 12
epochs, with batches of four distributed across two NVIDIA RTX 4090
GPUs. Optimization used AdamW with an initial learning rate of 1 x
10\textsuperscript{-5} , a weight decay of 1 x 10\textsuperscript{-5},
cosine learning-rate decay, and gradient clipping at a norm of five.
Fresh 1--10\% Gaussian noise was supplied to the CNN at each epoch,
whereas the TorchSpin reconstruction was compared with the corresponding
clean EasySpin spectrum. The spectral objective combined
first-derivative cosine loss, integrated-spectrum cosine loss, and
peak-weighted mean-squared error with relative weights of 1.0, 0.25, and
0.25, respectively. The integrated term emphasized agreement in the
underlying absorption-like envelope, whereas the weighted term increased
the contribution of high-intensity spectral regions. Parameter labels
were excluded from both the refinement loss and checkpoint selection;
parameter errors were recorded only as diagnostic quantities. The
refined checkpoint was selected using the same spectral objective on a
fixed subset of 128 validation spectra, with the initial supervised
checkpoint retained as the epoch-zero comparison.

\subsection*{Parameter-label-free training}

A separate experiment tested whether TorchSpin could train the same
inverse model from random initialization without parameter labels or
clean spectral targets. An independently generated EasySpin library
contained 4,096 training, 512 validation, and 512 prospective test
spectra sampled over the same Cu(II) domain. Only the fixed noisy
training and validation spectra were made available during
parameter-label-free optimization; their generating parameters and
noise-free counterparts were excluded from the training files. Each
observed spectrum was normalized to unit \emph{L\textsubscript{2}} norm,
supplied to the CNN, reconstructed through differentiable TorchSpin
\emph{pepper}, and compared with the same noisy observation using the
derivative, integrated, and peak-weighted spectral objective described
above. Three independently initialized models were trained using
different seeds, AdamW with an initial learning rate of 1 x
10\textsuperscript{-4}, a weight decay of 1 x 10\textsuperscript{-5},
batches of four, gradient clipping at a norm of five, and cosine
learning-rate decay over a maximum of 24 epochs. Checkpoints were
selected by the complete spectral loss on the noisy validation spectra,
with early stopping after six epochs without improvement. A separate
supervised control used the same architecture and fixed noisy
observations but was trained against the normalized generating
parameters. All four checkpoints were frozen before the prospective test
archive was accessed.

\subsection*{Cross-engine library transfer}

A separate prospective experiment tested whether a CNN trained on
TorchSpin-generated spectra could analyze spectra generated
independently with EasySpin. TorchSpin and EasySpin were each used to
generate independent libraries containing 4,096 training and 512
validation spectra within the same Cu(II) parameter domain, and a new
512-spectrum EasySpin test library was generated after the protocol was
frozen. Three CNNs were trained per engine using paired initialization
seeds; models sharing a seed began from identical weights but were
trained only on the library produced by their assigned simulator. All
six models used the architecture and supervised parameter objective
described above and completed 160 epochs, corresponding to 2,560
optimizer updates and 655,360 training-example backward passes per
model. Checkpoints were selected solely by parameter error on the
corresponding engine-specific validation set, without early stopping,
spectral refinement, or test-based selection. During evaluation,
parameters predicted by both conditions were reconstructed using the
same TorchSpin decoder so that the paired comparison isolated the effect
of the training-library engine. Exact duplicate checks found no overlap
between the prospective test set and either training library at the
parameter or normalized-spectrum level, and all six checkpoints were
frozen before the test archive was accessed.

\subsection*{CNN evaluation}

Spectral reconstruction was evaluated against the corresponding
noise-free EasySpin spectrum for every held-out test observation. The
primary metric was cosine similarity between
unit-\emph{L\textsubscript{2}}-normalized first-derivative spectra. We
additionally calculated normalized root-mean-square error, the
10th-percentile and minimum cosine similarities, and the fraction of
spectra reaching a cosine similarity of at least 0.950. Because both
spectra were normalized before comparison, cosine similarity and
normalized root-mean-square error describe the same underlying spectral
deviation and were not treated as independent validation metrics.
Parameter recovery was evaluated using component-wise mean absolute
errors (MAE) in physical units and, where comparisons across parameters
were required, as a fraction of each parameter's complete prespecified
sampling range. Results were also stratified into six fixed noise
intervals spanning 1--10\% added noise. Boundary occupancy, output
variance, and disagreement among models trained from different random
seeds were examined to identify collapsed solutions or spectrally
similar reconstructions produced from different parameter estimates.

\subsection*{Inference and reconstruction timing}

Latency was measured using the final EasySpin-trained CNN.
Single-spectrum inference was evaluated with one CPU thread and on one
NVIDIA RTX 4090 using 500 timed repetitions after warm-up. GPU timing
was performed both with an already normalized, device-resident tensor
and from a host-resident NumPy spectrum, with the latter including
\emph{L\textsubscript{2}} normalization, host-to-device transfer, and
CNN inference. Batched GPU throughput was measured for 1, 32, 128, and
512 device-resident spectra using 200 timed repetitions after 30 warm-up
calls. CUDA was synchronized immediately before timing and after each
measured call. A complete 512-point TorchSpin \emph{pepper}
reconstruction from the CNN-predicted parameters was timed separately
using 20 repetitions after five warm-up calculations. CNN inference and
TorchSpin reconstruction are reported separately because the network
returns parameter estimates directly, whereas the forward simulation is
an optional subsequent step used to evaluate or visualize spectral
agreement.

\FloatBarrier
\section*{Acknowledgments}

AHF would like to acknowledge UC Davis for start-up funding and is
supported by the Department of Energy, Laboratory Directed Research and
Development program at SLAC National Accelerator Laboratory, under
contract DE-AC0276SF00515. The author would also like to acknowledge
Stefan Stoll and other contributors to the development of EasySpin that
enabled this work.

\FloatBarrier
\section*{Code and data availability}

TorchSpin source code, documentation, worked notebooks, and executable
examples are available at \url{https://github.com/follmerlab/torchspin}.
The reference-comparison and performance analyses used TorchSpin v0.2.2.

\FloatBarrier

% Supplementary Information is part of the same compiled document.
\clearpage
\setcounter{page}{1}
\setcounter{figure}{0}
\setcounter{table}{0}
\renewcommand{\thepage}{S\arabic{page}}
\renewcommand{\thefigure}{S\arabic{figure}}
\renewcommand{\thetable}{S\arabic{table}}
% Keep hyperlink destinations distinct after resetting the counters.
\renewcommand{\theHfigure}{S.\arabic{figure}}
\renewcommand{\theHtable}{S.\arabic{table}}
\setlength{\parindent}{0pt}
\pdfbookmark[0]{Supplementary Information}{supplementary-information}
% Included from main.tex; not a standalone document.

\begin{center}
{\LARGE\bfseries TorchSpin: Differentiable Electron Paramagnetic Resonance
Simulation in PyTorch\par}
\vspace{12pt}
Alec H. Follmer\textsuperscript{1,2,*}\par
\vspace{6pt}
{\small \emph{\textsuperscript{1} Department of Chemistry, University of
California, Davis, Davis, California 95616, United States}\par}
{\small \emph{\textsuperscript{2} Linac Coherent Light Source, SLAC National
Accelerator Laboratory, Menlo Park, California 94025, United States}\par}
{\small \textsuperscript{*}Corresponding author:
\href{mailto:ahfollmer@ucdavis.edu}{\nolinkurl{ahfollmer@ucdavis.edu}}\par}
\vspace{12pt}
{\large\bfseries Supplementary Information\par}
\end{center}

\section*{Table of Contents}
\begin{figure}[H]
\centering
\includegraphics[width=3.10162in,height=0.64\textheight,keepaspectratio]{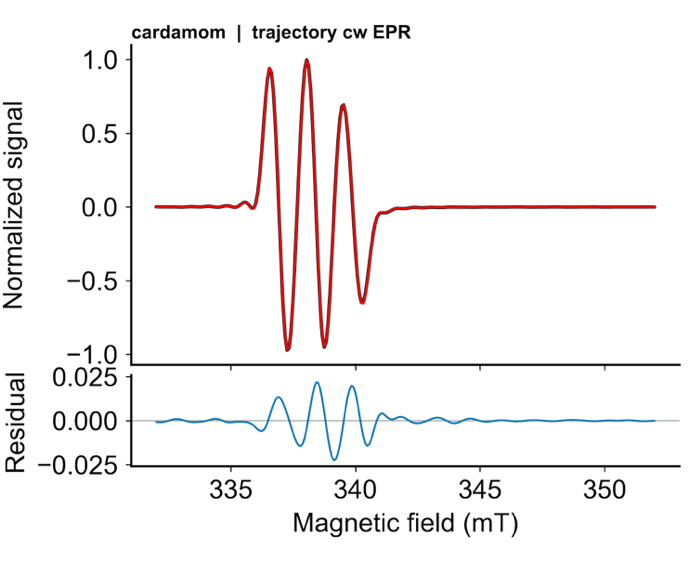}
\caption{Comparison of trajectory-based cw EPR simulations generated by TorchSpin
and EasySpin. TorchSpin (red) and EasySpin (black) spectra were
calculated using cardamom with the same spin-system and experimental
parameters. Blue shows the residual (TorchSpin − EasySpin). Because the
calculations use independently sampled rotational trajectories, the
residual includes finite-sampling variation. Each TorchSpin spectrum was
multiplied by a single amplitude factor to account for differences in
absolute-intensity convention.}
\label{fig:S1}
\end{figure}

\begin{figure}[H]
\centering
\includegraphics[width=6.50000in,height=0.64\textheight,keepaspectratio]{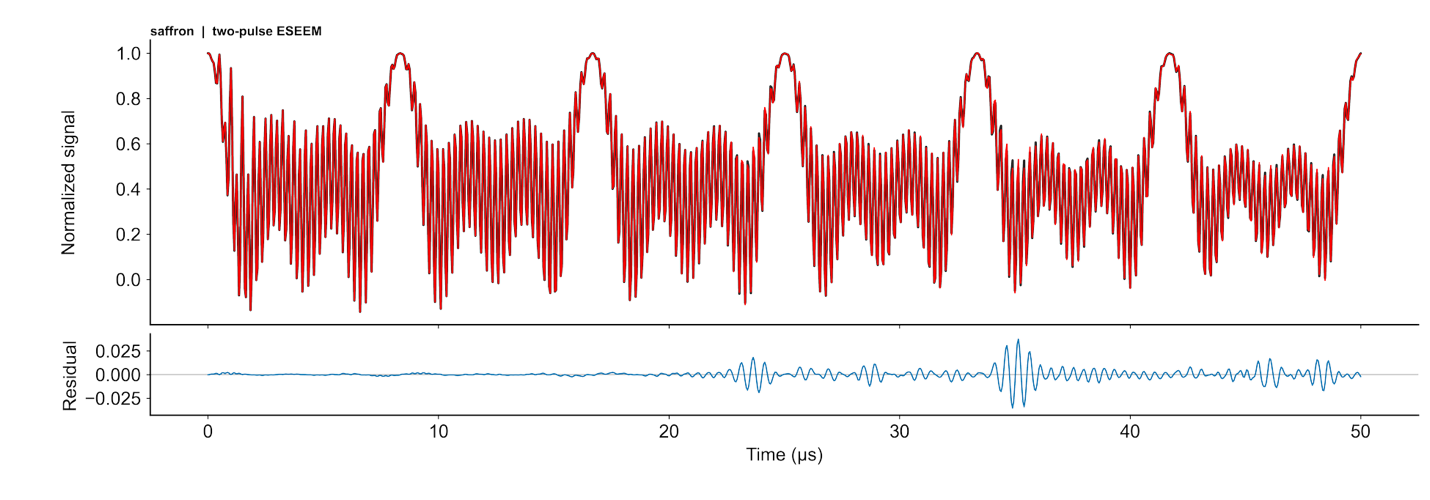}
\caption{Comparison of time-domain two-pulse ESEEM simulations generated by
TorchSpin and EasySpin. The real components of the TorchSpin (red) and
EasySpin (black) echo signals were calculated using saffron with
equivalent spin-system, pulse-sequence, and orientation-grid parameters.
Blue shows the residuals (TorchSpin − EasySpin). A single amplitude
factor was applied to the complete TorchSpin trace to account for
different absolute-intensity conventions.}
\label{fig:S2}
\end{figure}

\begin{figure}[H]
\centering
\includegraphics[width=4.34722in,height=0.64\textheight,keepaspectratio]{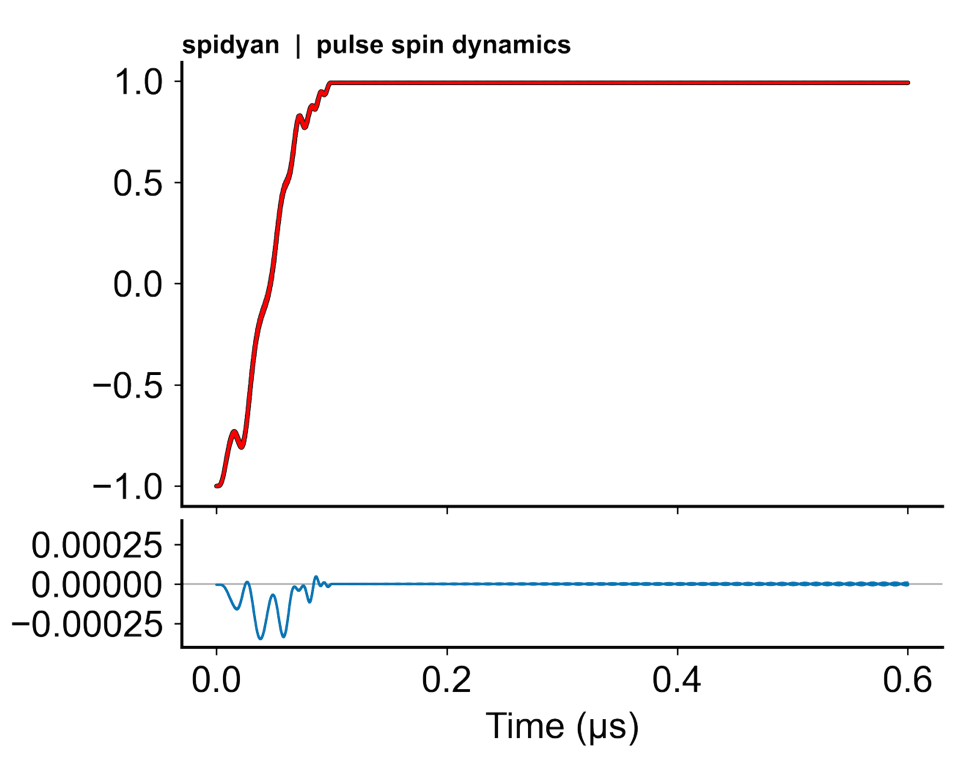}
\caption{Comparison of pulse-driven spin dynamics calculated by TorchSpin and
EasySpin. The real-valued electron \emph{z}-magnetization trajectories
generated by TorchSpin (red) and EasySpin (black) are shown during a
frequency-swept pulse and the subsequent free-evolution period. Blue
shows the pointwise residual (TorchSpin − EasySpin). A single amplitude
factor was applied to the complete TorchSpin trajectory to account for
differences in signal convention.}
\label{fig:S3}
\end{figure}

\begin{figure}[H]
\centering
\includegraphics[width=4.34722in,height=0.64\textheight,keepaspectratio]{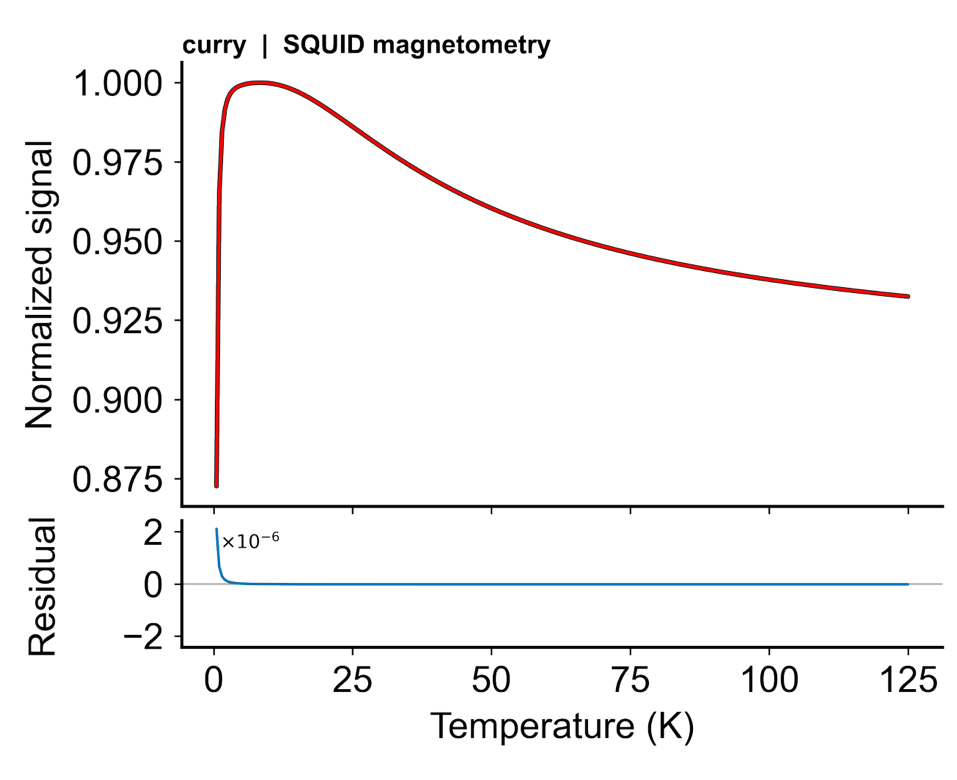}
\caption{Comparison of temperature-dependent effective magnetic moments
calculated by TorchSpin and EasySpin. The normalized effective magnetic
moments obtained with TorchSpin (red) and EasySpin (black) are shown as
a function of temperature for an exchange-coupled Gd(III)--Cu(II) spin
system. Blue shows the pointwise residual (TorchSpin − EasySpin),
magnified by 10\textsuperscript{6} for visibility. A single amplitude
factor was applied to the complete TorchSpin trace before normalization.}
\label{fig:S4}
\end{figure}

\begin{figure}[H]
\centering
\includegraphics[width=5.11608in,height=0.64\textheight,keepaspectratio]{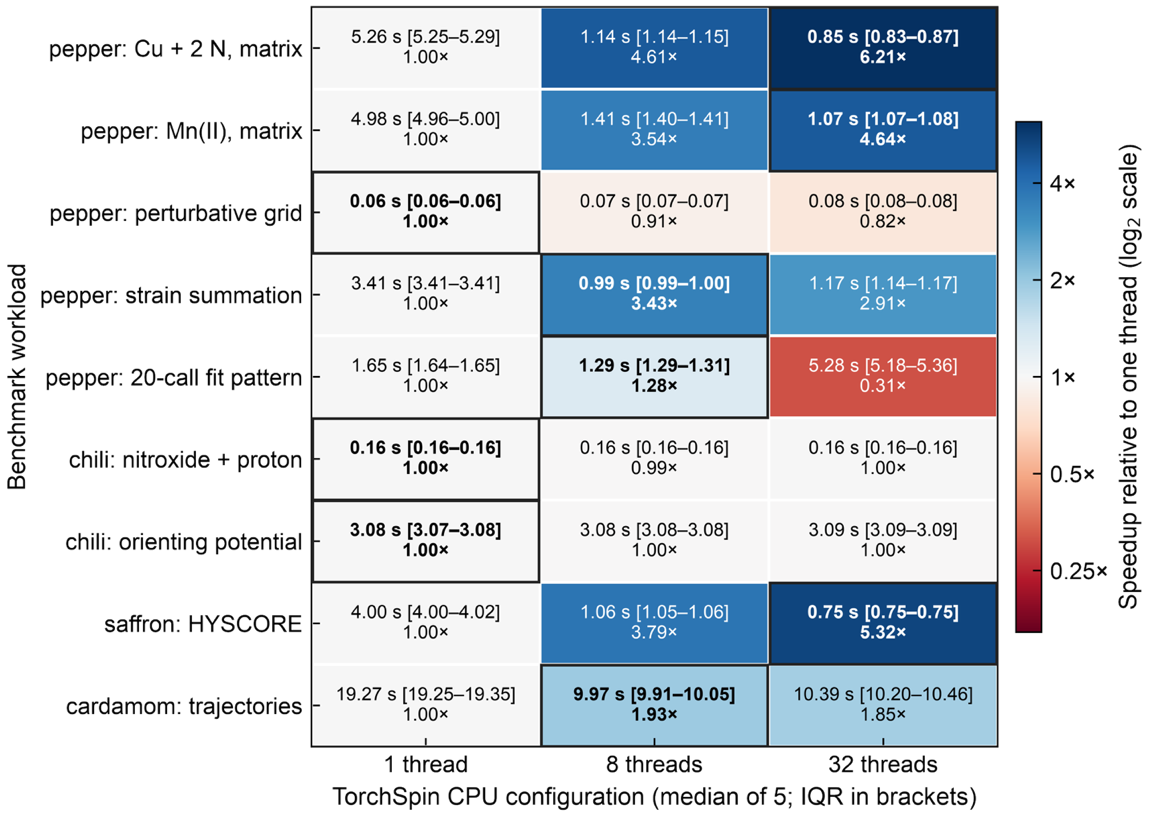}
\caption{CPU thread scaling of representative TorchSpin simulation workloads.
Median wall times were measured using 1, 8, and 32 CPU threads on a
local workstation. Each cell reports the median execution time with its
IQR in brackets and the speedup relative to the corresponding one-thread
calculation. Cell color encodes the speedup on a log\textsubscript{2}
scale: blue indicates faster execution, red indicates slower execution,
and white indicates little or no change. The black outline and bold text
identify the selected CPU configuration, defined as the smallest thread
count whose timing interquartile range overlapped that of the
configuration with the lowest median wall time. Each condition comprised
one untimed warm-up followed by five timed replicates; quartiles were
calculated as the linearly interpolated 25th and 75th percentiles.}
\label{fig:S5}
\end{figure}

\begin{figure}[H]
\centering
\includegraphics[width=4.85766in,height=0.64\textheight,keepaspectratio]{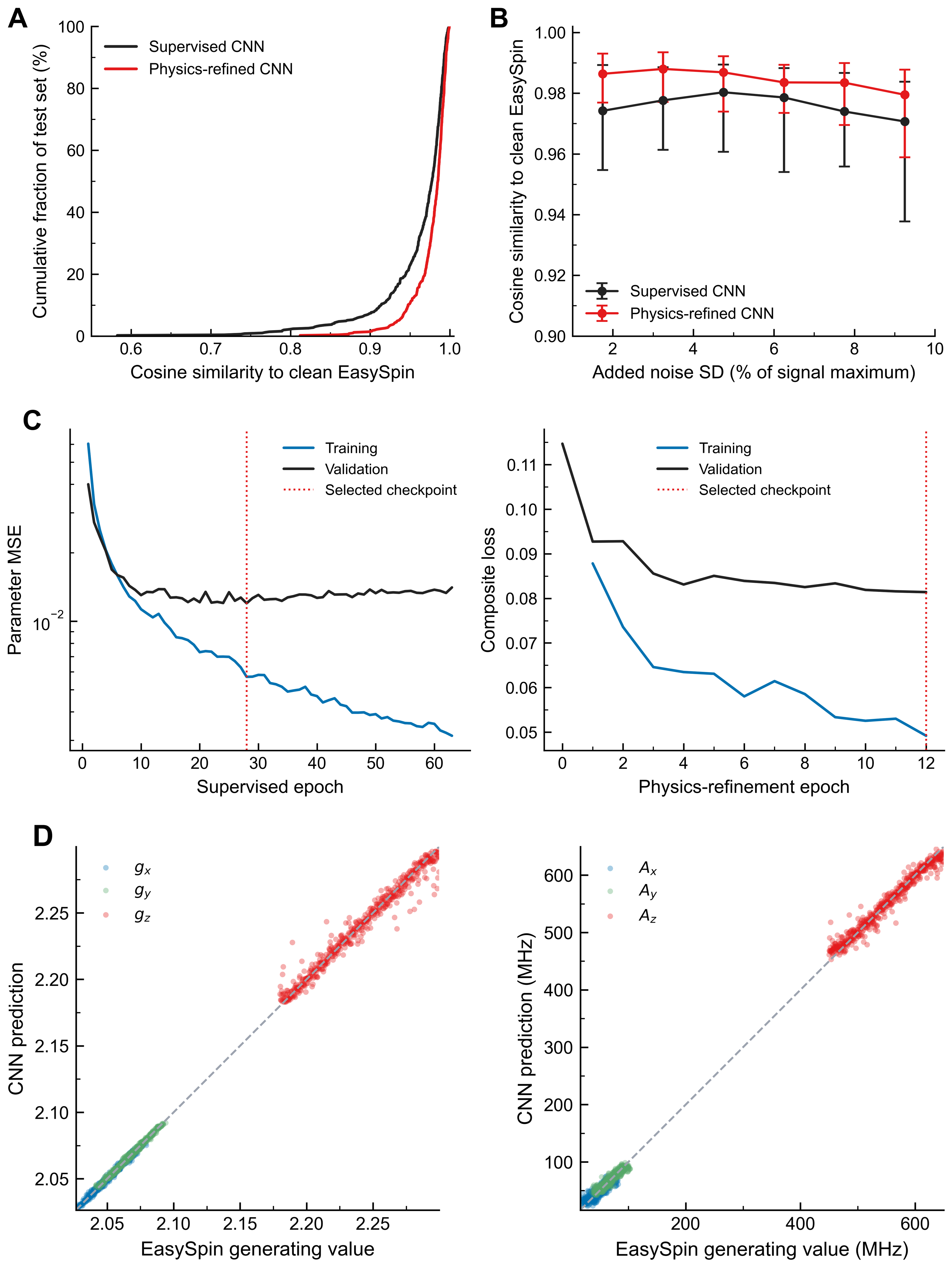}
\caption{Analysis of differentiable spectral refinement after supervised
initialization. (\textbf{A}) Cumulative distribution of cosine
similarity to the clean EasySpin test spectrum for the supervised and
physics-refined CNNs. (\textbf{B}) Median cosine similarity within six
predefined noise intervals; error bars show the interquartile range. The
bin counts from lowest to highest noise were 82, 97, 84, 83, 79, and 87
spectra. (\textbf{C}) Training and validation histories for supervised
parameter regression (\emph{left}) and parameter-label-free spectral
refinement (\emph{right}); dotted lines mark the selected checkpoints.
(\textbf{D}) Predicted and generating \emph{g} and hyperfine components
for the refined model.}
\label{fig:S6}
\end{figure}

\begin{figure}[H]
\centering
\includegraphics[trim=0.000000bp 1444.641450bp 0.000000bp 0.000000bp,clip,width=6.45000in,height=0.64\textheight,keepaspectratio]{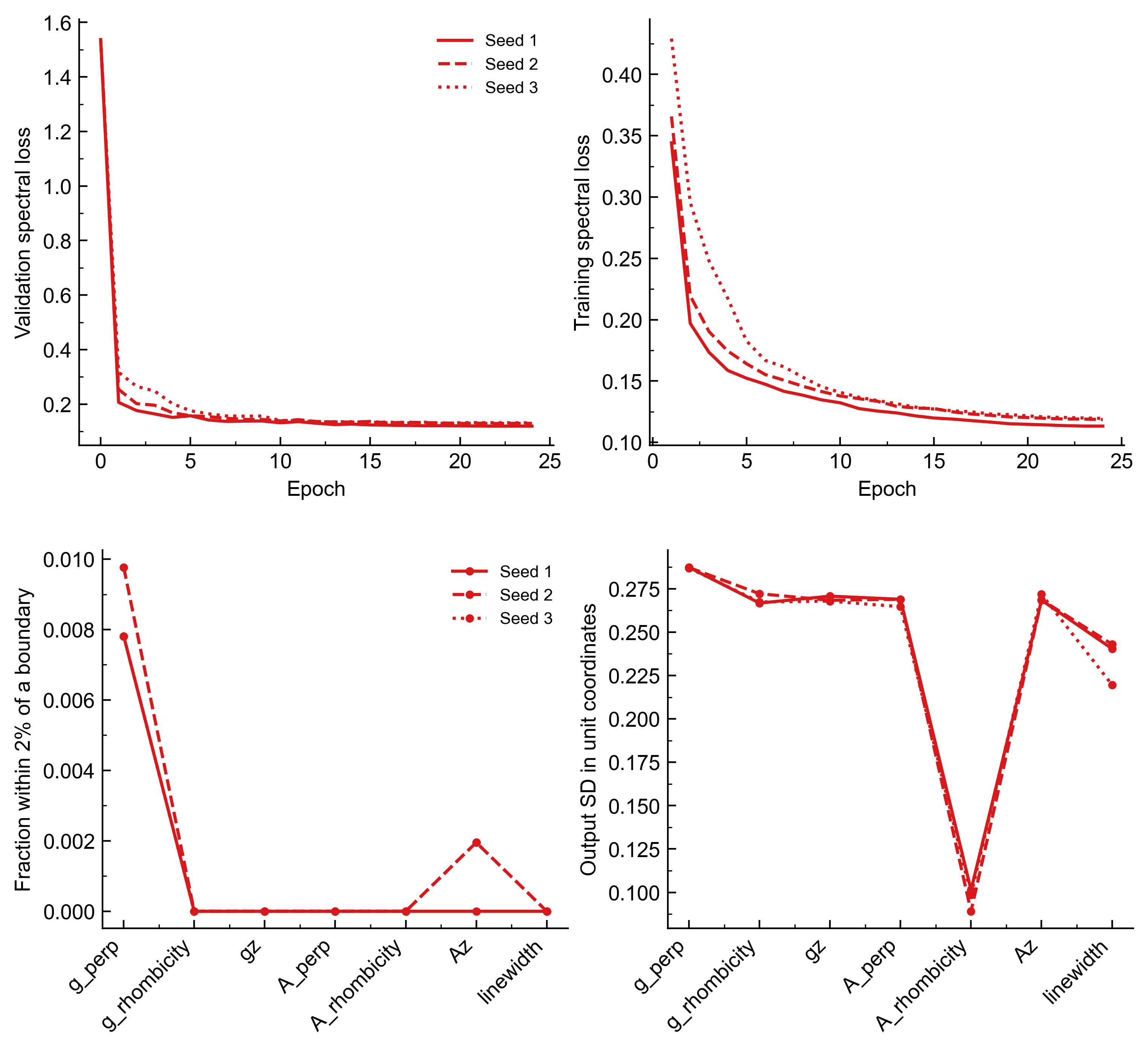}
\caption{Optimization histories for the three parameter-label-free CNNs.
Validation spectral loss and training spectral loss across the 24-epoch
schedules. Line styles distinguish the three initialization seeds.}
\label{fig:S7}
\end{figure}

\begin{figure}[H]
\centering
\includegraphics[width=5.20909in,height=0.64\textheight,keepaspectratio]{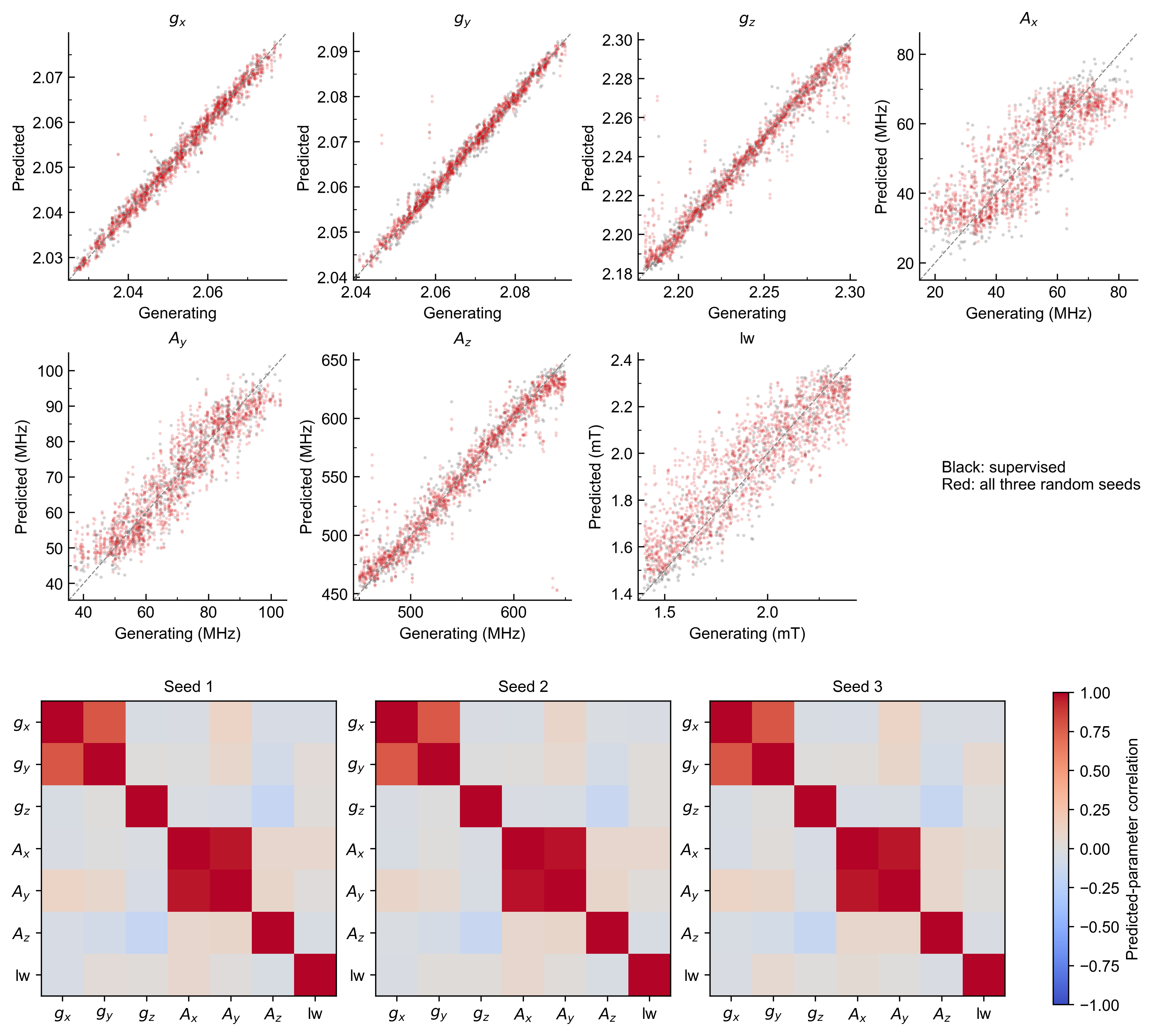}
\caption{Parameter recovery and predicted-parameter correlations for training
without labels. \emph{Top}: generating versus predicted physical
parameters for the supervised control (black) and all three label-free
seeds (red); the dashed line denotes equality. \emph{Bottom}:
correlation matrices of the seven predicted parameters for each
label-free seed.}
\label{fig:S8}
\end{figure}

\begin{figure}[H]
\centering
\includegraphics[width=6.45000in,height=0.64\textheight,keepaspectratio]{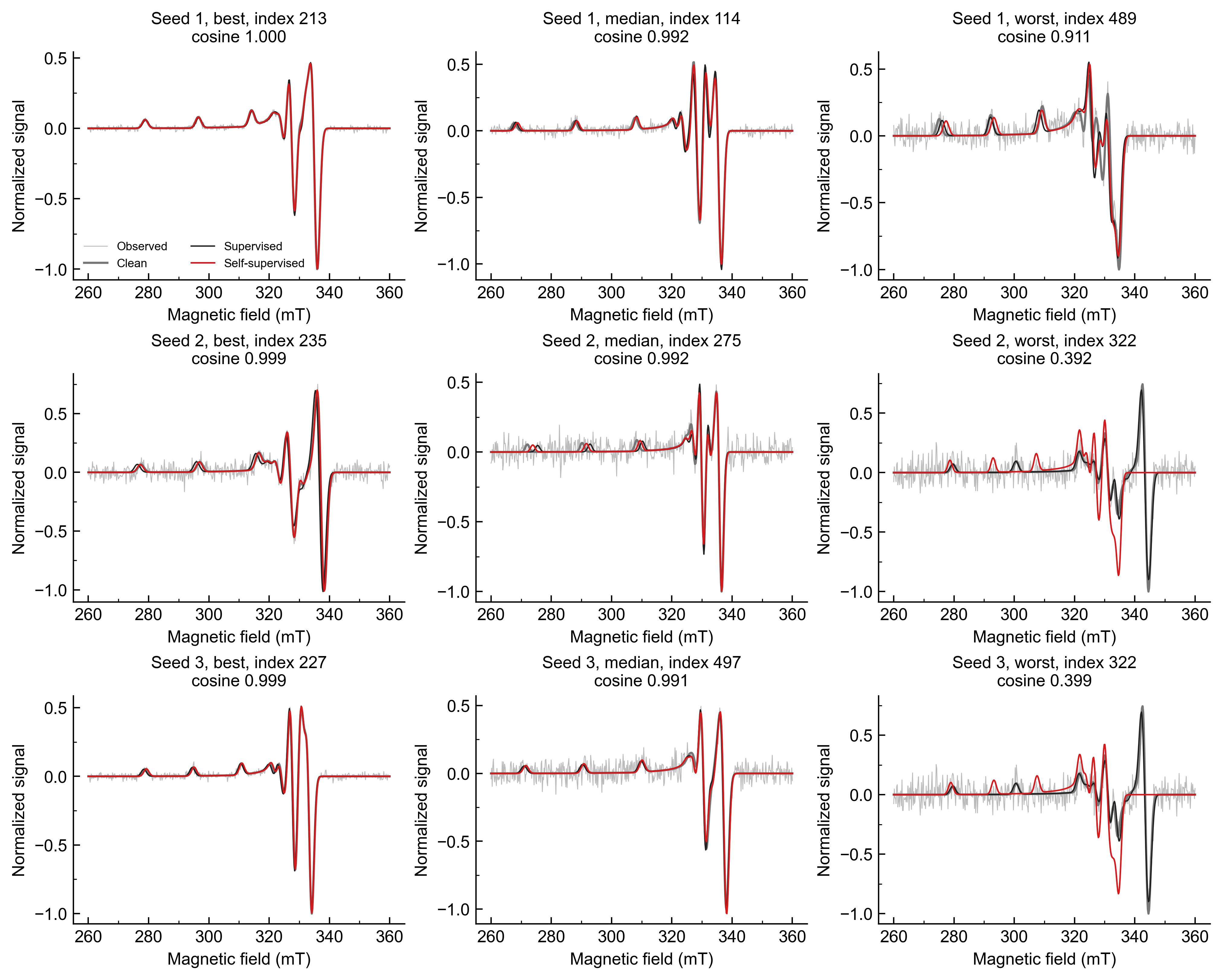}
\caption{Representative parameter-label-free reconstructions across the
performance distribution. Best, median-ranked, and worst examples are
shown for each initialization seed. Gray traces are fixed noisy
observations, black traces are clean EasySpin spectra, and red traces
are TorchSpin reconstructions from the label-free CNN predictions. Test
indices and cosine similarities are given above each panel.}
\label{fig:S9}
\end{figure}

\begin{figure}[H]
\centering
\includegraphics[width=4.24043in,height=0.64\textheight,keepaspectratio]{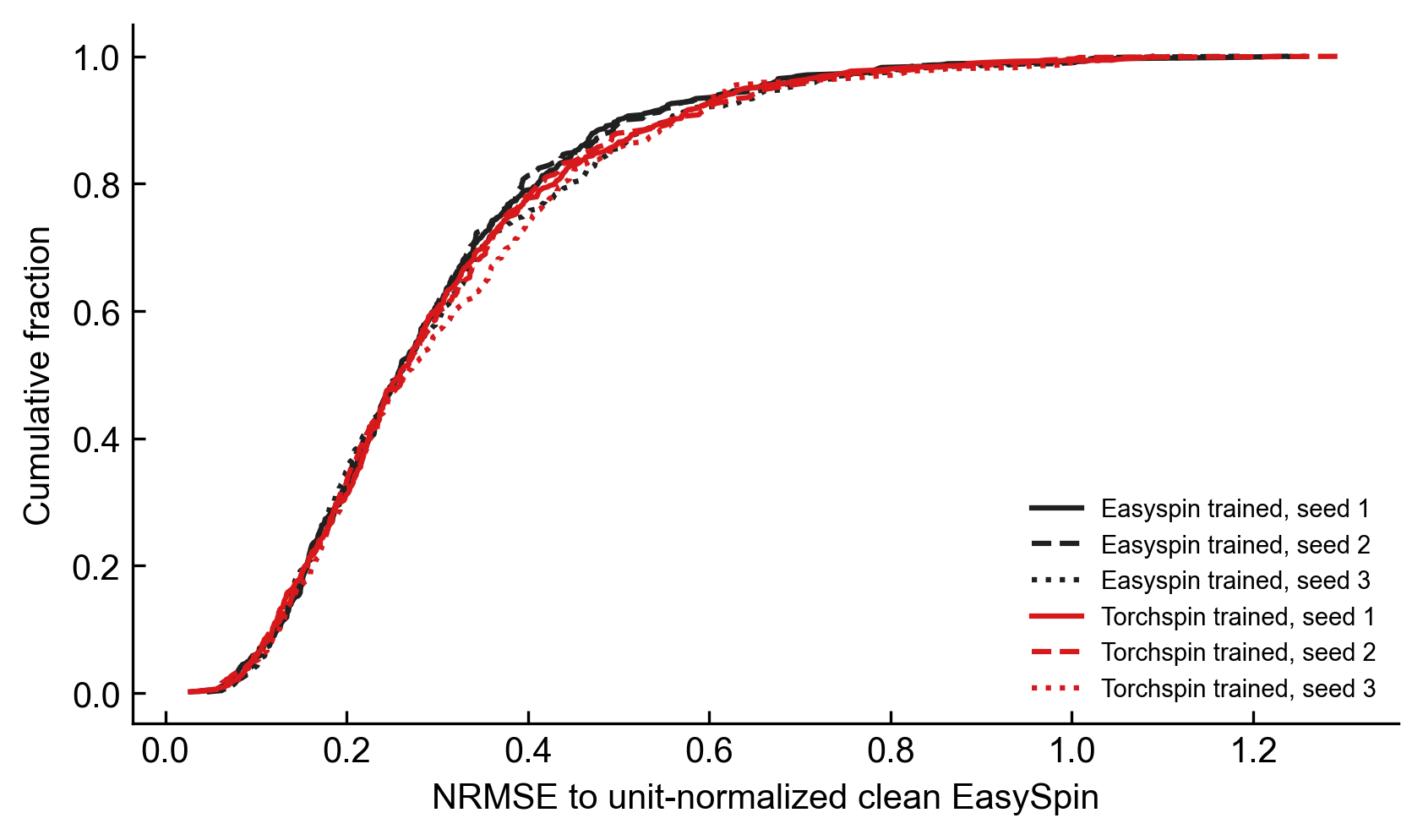}
\caption{Cumulative distributions of reconstruction NRMSE for CNNs trained on
labeled EasySpin or TorchSpin libraries and evaluated on the same
prospective EasySpin test set. Line styles identify the three paired
initialization seeds. All predictions were decoded with TorchSpin.}
\label{fig:S10}
\end{figure}

\begin{figure}[H]
\centering
\includegraphics[width=0.485\linewidth]{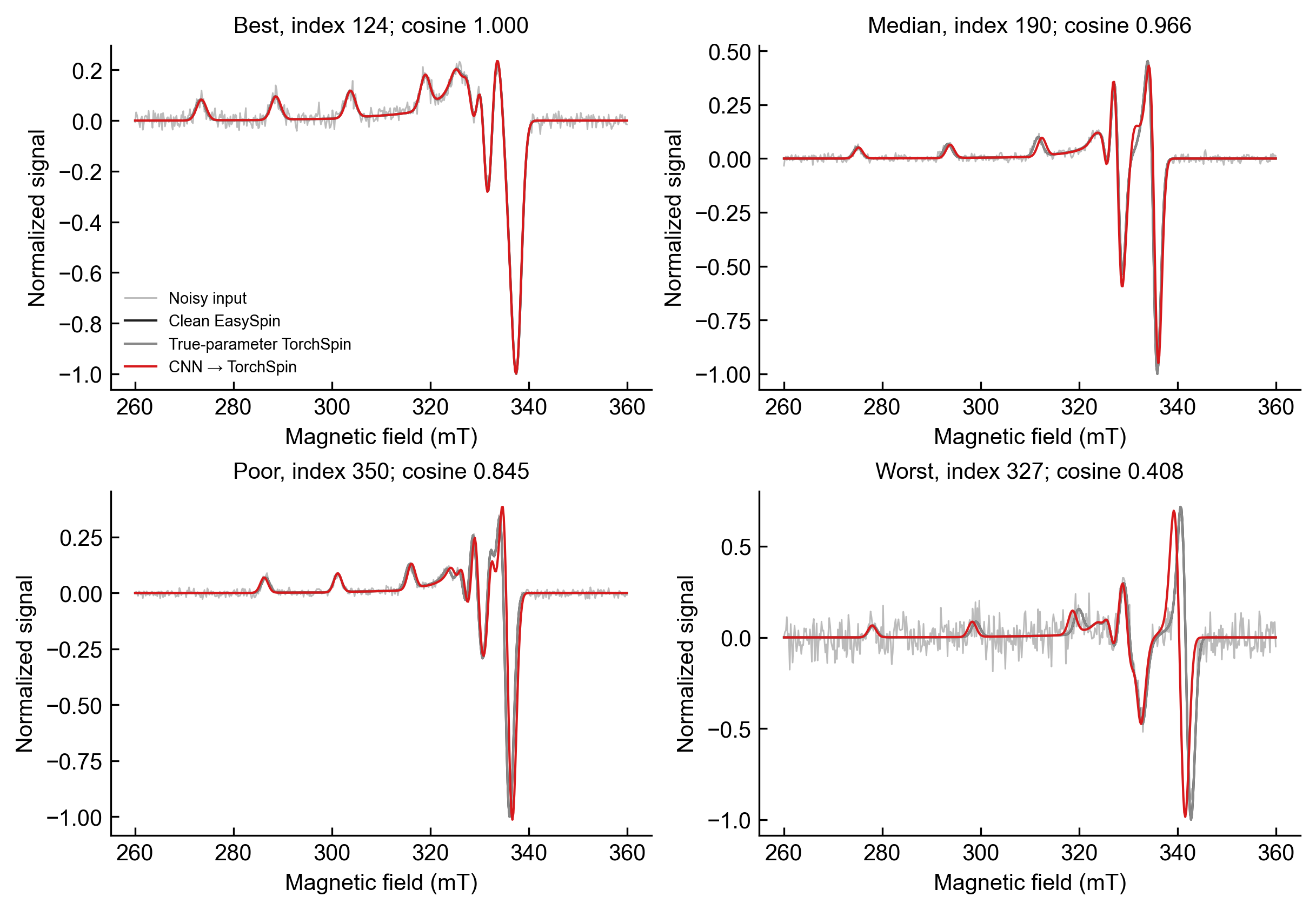}\hfill
\includegraphics[width=0.485\linewidth]{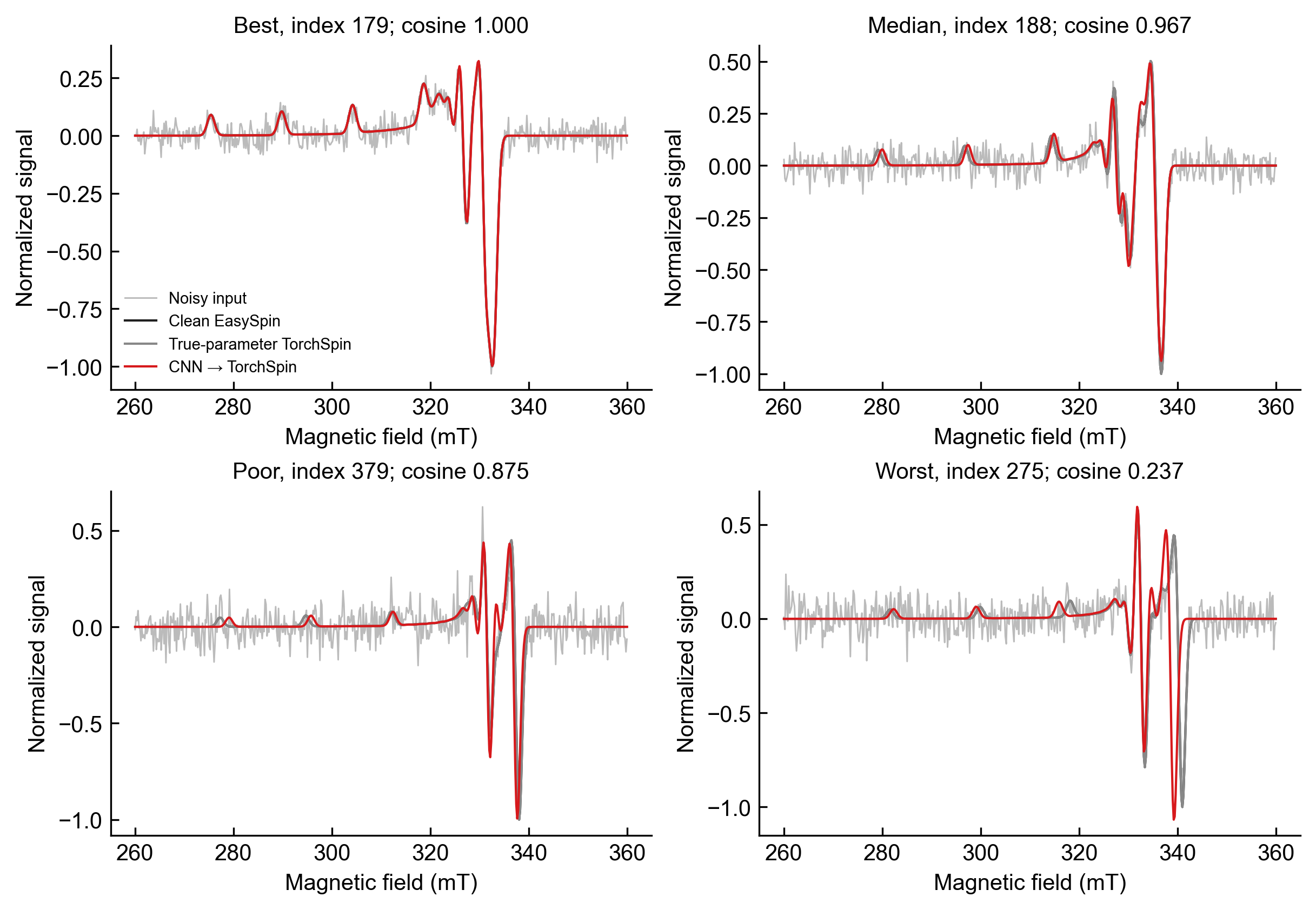}
\caption{Ranked prospective reconstructions for paired initialization seed 1.
\emph{Left}: CNN trained on the TorchSpin library. \emph{Right}: CNN
trained on the EasySpin library. For each condition, the best,
median-ranked, poor, and worst cases were selected by cosine similarity
to the clean EasySpin test spectrum. Gray traces are noisy inputs, black
traces are clean EasySpin references, and red traces are TorchSpin
reconstructions.}
\label{fig:S11}
\end{figure}

\begin{figure}[H]
\centering
\includegraphics[width=0.485\linewidth]{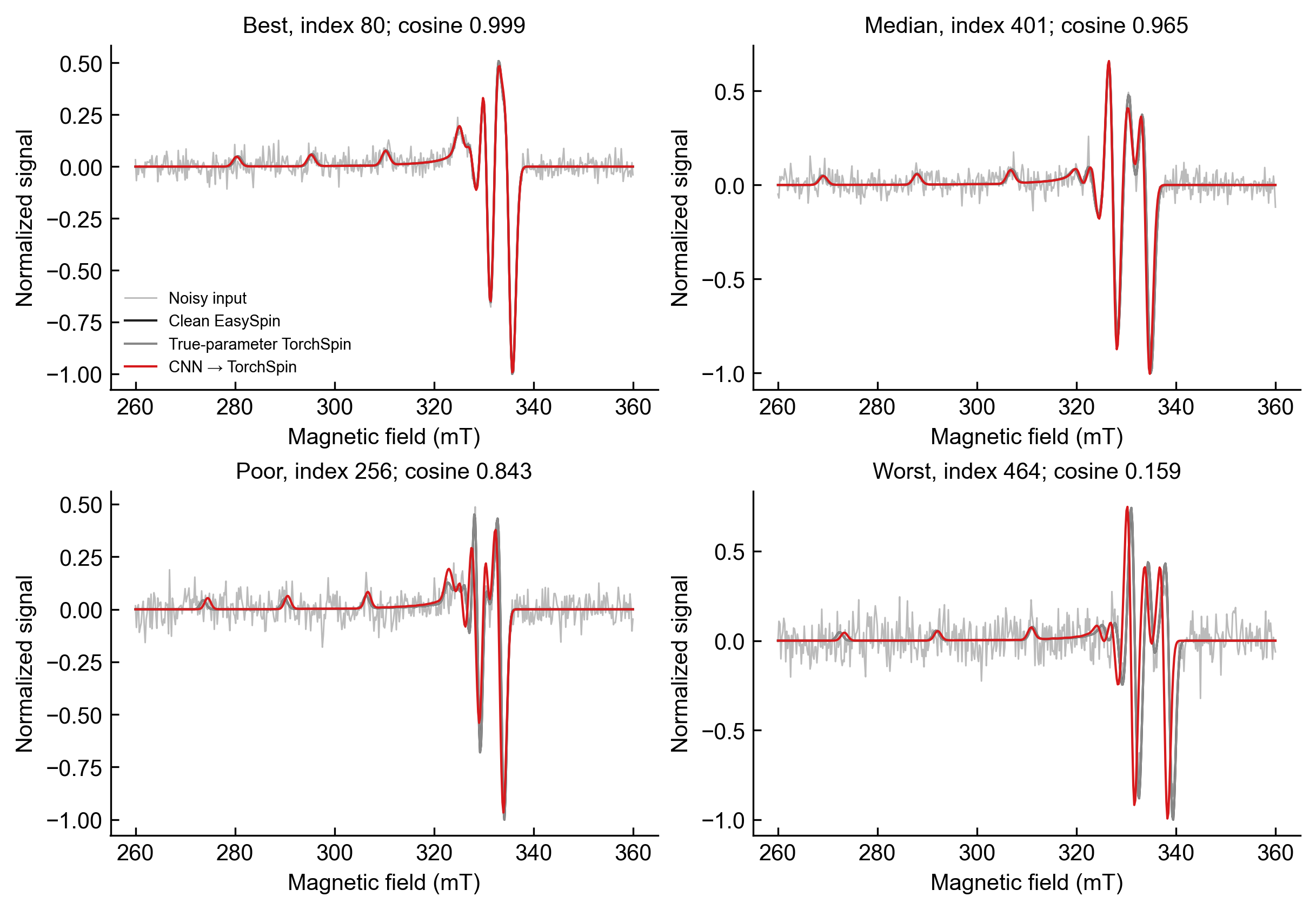}\hfill
\includegraphics[width=0.485\linewidth]{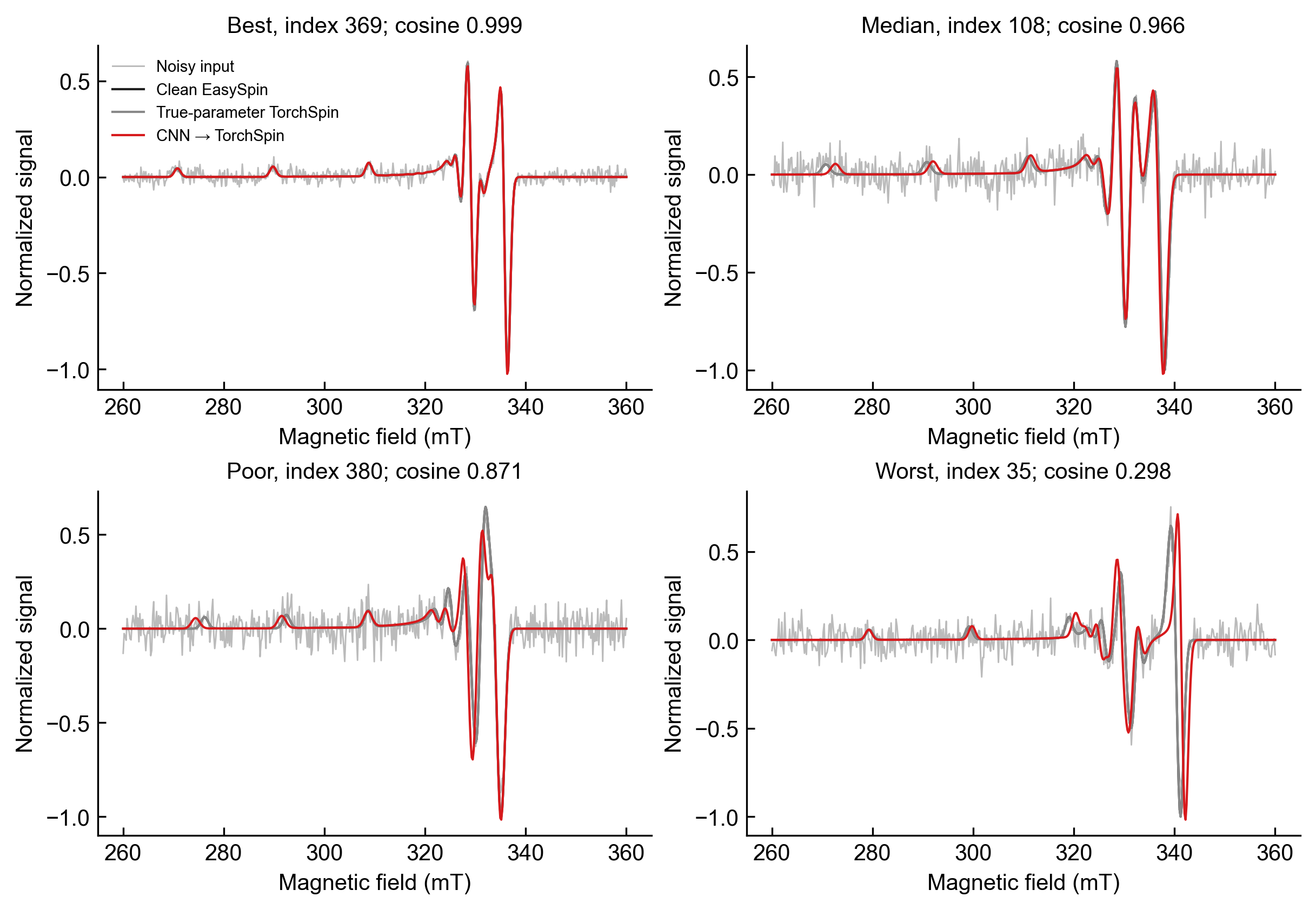}
\caption{Ranked prospective reconstructions for paired initialization seed 2.
\emph{Left}: CNN trained on the TorchSpin library. \emph{Right}: CNN
trained on the EasySpin library. For each condition, the best,
median-ranked, poor, and worst cases were selected by cosine similarity
to the clean EasySpin test spectrum. Gray traces are noisy inputs, black
traces are clean EasySpin references, and red traces are TorchSpin
reconstructions.}
\label{fig:S12}
\end{figure}

\begin{figure}[H]
\centering
\includegraphics[width=0.485\linewidth]{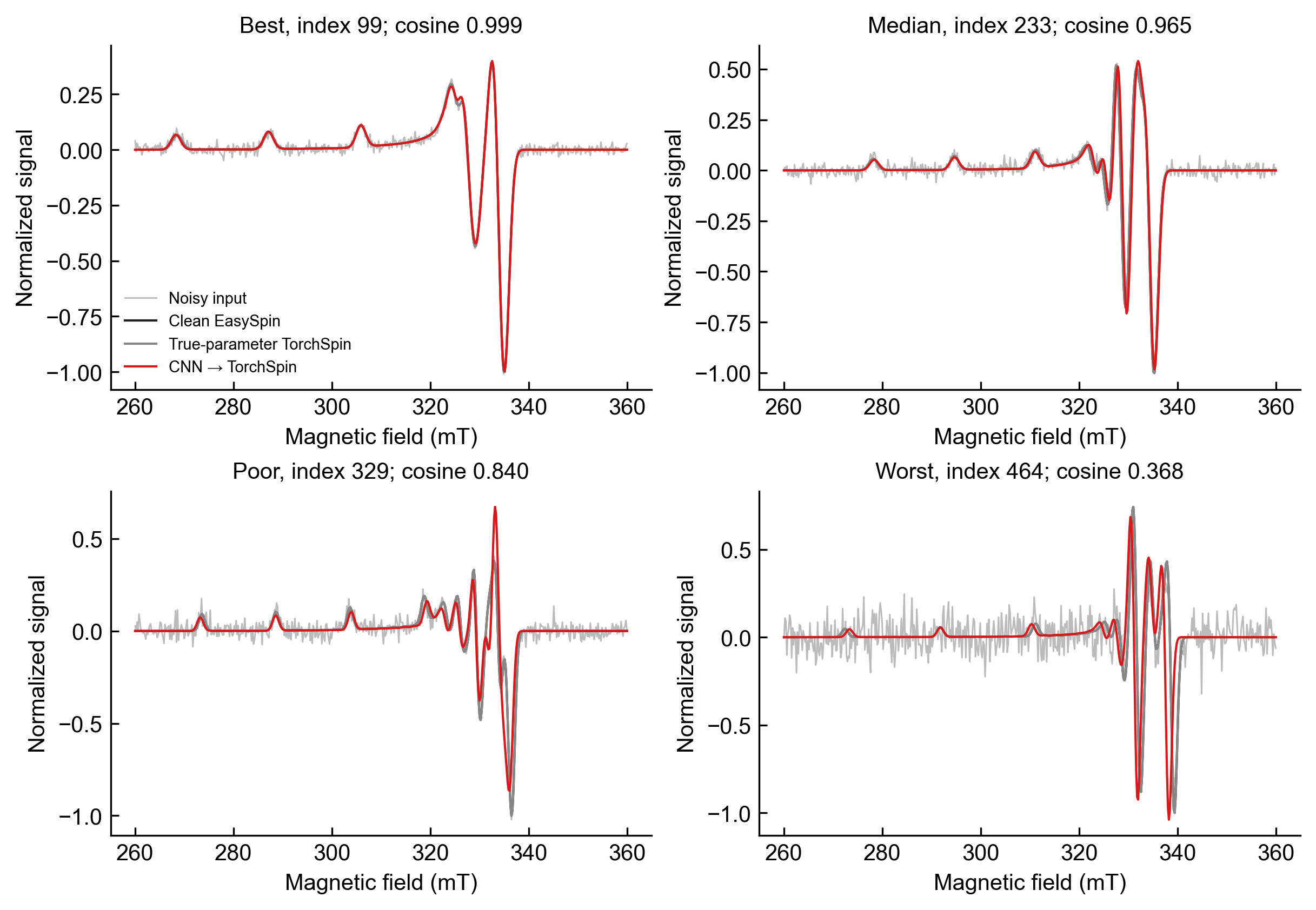}\hfill
\includegraphics[width=0.485\linewidth]{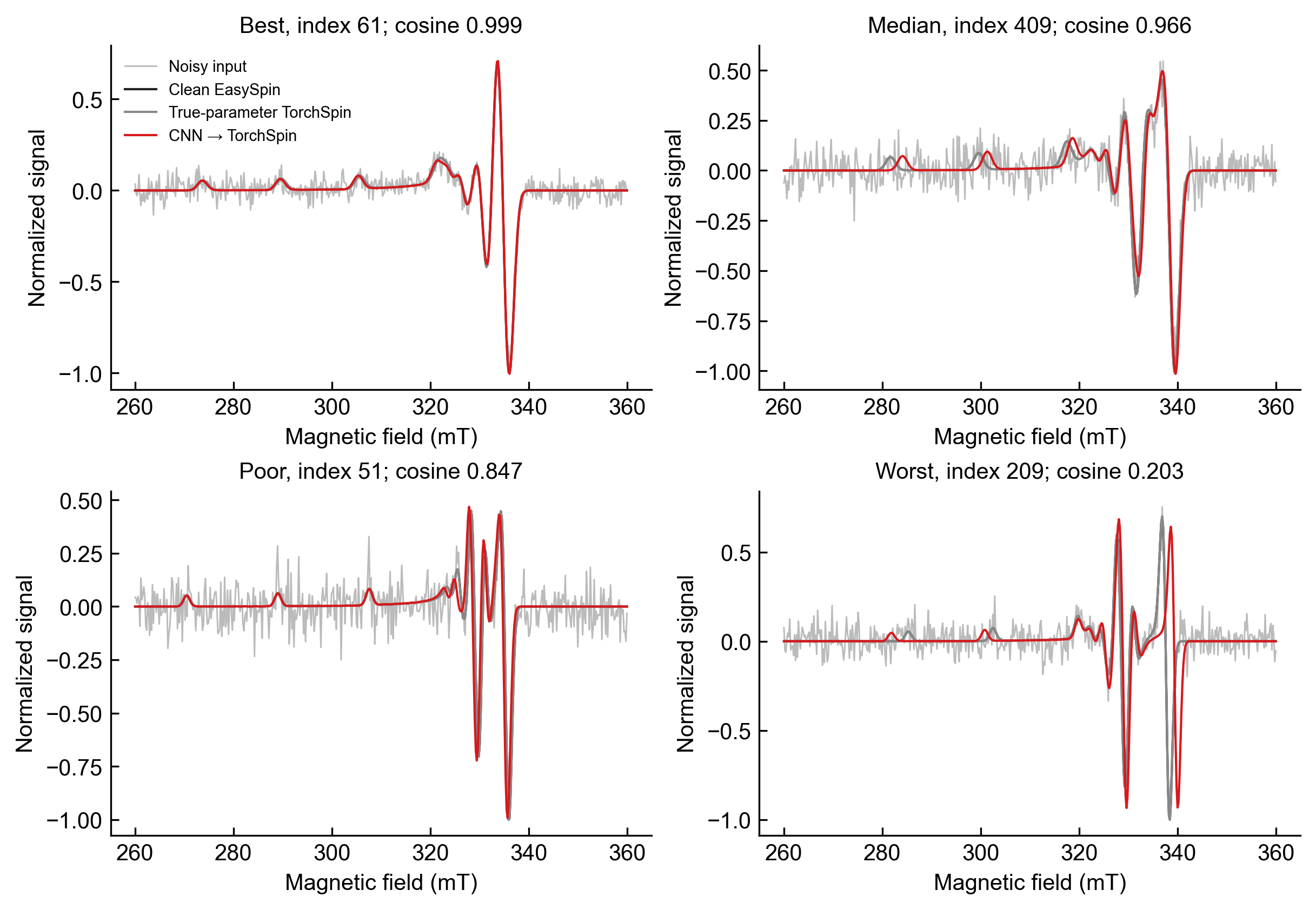}
\caption{Ranked prospective reconstructions for paired initialization seed 3.
\emph{Left}: CNN trained on the TorchSpin library. \emph{Right}: CNN
trained on the EasySpin library. For each condition, the best,
median-ranked, poor, and worst cases were selected by cosine similarity
to the clean EasySpin test spectrum. Gray traces are noisy inputs, black
traces are clean EasySpin references, and red traces are TorchSpin
reconstructions.}
\label{fig:S13}
\end{figure}

\begingroup\small
\setlength{\parskip}{3pt}
\noindent\hyperref[fig:S1]{Figure S1. Comparison of cw EPR simulations generated by TorchSpin and EasySpin}\dotfill\pageref{fig:S1}\par
\noindent\hyperref[fig:S2]{Figure S2. Comparison of two-pulse ESEEM simulations generated by TorchSpin and EasySpin}\dotfill\pageref{fig:S2}\par
\noindent\hyperref[fig:S3]{Figure S3. Comparison of pulse-driven spin dynamics calculated by TorchSpin and EasySpin}\dotfill\pageref{fig:S3}\par
\noindent\hyperref[fig:S4]{Figure S4. Comparison of magnetic moments calculated by TorchSpin and EasySpin}\dotfill\pageref{fig:S4}\par
\noindent\hyperref[fig:S5]{Figure S5. CPU thread scaling of representative TorchSpin simulation workloads}\dotfill\pageref{fig:S5}\par
\noindent\hyperref[tab:S1]{Table S1. Matched-host EasySpin and TorchSpin performance measurements}\dotfill\pageref{tab:S1}\par
\noindent\hyperref[tab:S2]{Table S2. Bounded coordinates used to generate the Cu(II) spectral libraries}\dotfill\pageref{tab:S2}\par
\noindent\hyperref[fig:S6]{Figure S6. Supplementary analysis of differentiable spectral refinement after supervised initialization}\dotfill\pageref{fig:S6}\par
\noindent\hyperref[tab:S3]{Table S3. Test-set effects of parameter-label-free spectral refinement after supervised initialization}\dotfill\pageref{tab:S3}\par
\noindent\hyperref[tab:S4]{Table S4. Component-wise parameter errors from parameter-label-free spectral refinement}\dotfill\pageref{tab:S4}\par
\noindent\hyperref[fig:S7]{Figure S7. Optimization histories for parameter-label-free CNN training from random initialization}\dotfill\pageref{fig:S7}\par
\noindent\hyperref[fig:S8]{Figure S8. Parameter recovery and parameter correlations for training without parameter labels}\dotfill\pageref{fig:S8}\par
\noindent\hyperref[fig:S9]{Figure S9. Representative parameter-label-free spectral reconstructions}\dotfill\pageref{fig:S9}\par
\noindent\hyperref[tab:S5]{Table S5. Prospective test results for parameter-label-free training from random initialization}\dotfill\pageref{tab:S5}\par
\noindent\hyperref[fig:S10]{Figure S10. Reconstruction-error distributions for CNNs trained on EasySpin and TorchSpin libraries}\dotfill\pageref{fig:S10}\par
\noindent\hyperref[fig:S11]{Figure S11. Ranked prospective reconstructions for paired CNN initialization seed 1}\dotfill\pageref{fig:S11}\par
\noindent\hyperref[fig:S12]{Figure S12. Ranked prospective reconstructions for paired CNN initialization seed 2}\dotfill\pageref{fig:S12}\par
\noindent\hyperref[fig:S13]{Figure S13. Ranked prospective reconstructions for paired CNN initialization seed 3}\dotfill\pageref{fig:S13}\par
\noindent\hyperref[tab:S6]{Table S6. Prospective cross-engine comparison across three paired CNN initialization seeds}\dotfill\pageref{tab:S6}\par
\noindent\hyperref[tab:S7]{Table S7. Component-wise parameter errors in the cross-engine training-library comparison}\dotfill\pageref{tab:S7}\par
\noindent\hyperref[tab:S8]{Table S8. CNN inference and TorchSpin reconstruction latency}\dotfill\pageref{tab:S8}\par
\endgroup
\clearpage

\textbf{Simulation details.} The spin system was composed of an S=1/2
system coupled to one \textsuperscript{14}N nucleus, with principal
values \emph{g} = {[}2.008, 2.006, 2.003{]} and \emph{A} = {[}20, 20,
85{]} MHz. Rotational diffusion was modeled with a correlation time of 1
ns. The field-swept absorption spectrum was calculated at a microwave
frequency of 9.5 GHz over 332--352 mT using 256 points. Each calculation
used 100 trajectories of 500 steps, with spin and spatial propagation
time steps of 0.1 ns. The EasySpin reference used a fixed random seed,
whereas the current TorchSpin interface does not expose the trajectory
seed.

\textbf{Simulation details.} The spin system was defined as an electron
(S = 1/2) coupled to one \textsuperscript{14}N nucleus, with an
isotropic \emph{g} value of 2.0023 and an isotropic hyperfine coupling
of 2.000 MHz. The \textsuperscript{14}N quadrupole interaction was
defined by \(e^{2}qQ/h = 0.400\) MHz and an asymmetry parameter
\(\eta\ \)= 0.6. A two-pulse ESEEM sequence was simulated at a magnetic
field of 324.9 mT with an initial delay of 0.001 µs. The evolution
period was sampled at 0.050 µs intervals using 1001 points, and the
powder average used an orientation-grid size of 20.

\textbf{Simulation details.} The spin system contained an S = 1/2 system
with an isotropic \emph{g}-value of 2.00232 coupled to one
\textsuperscript{1}H nucleus with hyperfine principal values \emph{A} =
{[}8, 45, 45{]} MHz. The simulation was performed at 1195 mT with a
microwave carrier frequency of 33.5 GHz. The sequence consisted of a
0.100 µs linear chirp pulse with quarter-sine-smoothed edges, a rise
time of 0.015 µs, a frequency sweep from −100 to +100 MHz relative to
the carrier, and a nominal flip angle of \emph{π}, followed by 0.5 µs of
free evolution. \emph{S\textsubscript{z}} was detected throughout both
sequence elements at zero detection frequency and phase. Propagation
used an integration time step of 0.0001 µs and a 32 GHz simulation
frame.

\textbf{Simulation details.} The model contained an S = 7/2 Gd(III)
center and an S = 1/2 Cu(II) center, each with an isotropic
\emph{g}-value of 2.0023. The spins were coupled through the isotropic
exchange Hamiltonian (\(H_{ex} = - 2J\mathbf{S}1 \cdot \mathbf{S}2\)),
with \emph{J} = 5 cm\textsuperscript{-1}, represented in the simulation
by an electron-electron coupling of -300 GHz. Molar susceptibility was
calculated at 100 mT from 0.5 to 125 K in 0.5 K increments using an
orientation-grid size of 10 and a field increment of 0.01 mT. The
TorchSpin susceptibility was converted to the effective magnetic moment
using
(\(\mu_{eff} = \left\lbrack 3k_{B}\chi_{mol}T/\left( \mu_{0}N_{A}\mu_{B}^{2} \right) \right\rbrack^{1/2}\))
before comparison with the EasySpin result.

\FloatBarrier
\section*{Performance benchmark design}

The benchmark suite was designed to sample the principal computational
workloads encountered across the current TorchSpin simulation framework
rather than to exhaustively test every available option. The selected
calculations probe matrix diagonalization, perturbative resonance
calculations, strain summation, repeated forward simulations,
stochastic-Liouville calculations, multidimensional pulse EPR, and
trajectory-based spin dynamics.

All benchmarks were performed on the same workstation using a fixed
allocation of 32 physical CPU cores. EasySpin calculations were
performed using version 6.1.1 in MATLAB R2026a, whereas TorchSpin was
evaluated with 1, 8, and 32 CPU threads. CUDA timings were collected
only for workloads with an implemented accelerator path. These
calculations are labeled ``TorchSpin CUDA mode'' because they combine
GPU-resident tensor operations with CPU-side orchestration and therefore
should not be interpreted as entirely GPU-resident implementations. The
two \emph{chili} workloads and the \emph{cardamom} workload currently
lack CUDA execution paths and are consequently reported as not
applicable (N/A).

\subsection*{pepper: Cu(II) with two nitrogen nuclei}

The first matrix-based \emph{pepper} workload represents an S= 1/2
Cu(II) center coupled to one \textsuperscript{63}Cu nucleus and two
\textsuperscript{14}N nuclei. The \emph{g} tensor was set to {[}2.05,
2.05, 2.25{]}, and the \emph{A} tensors were {[}30, 30, 500{]} MHz for
copper and {[}40, 40, 40{]} MHz for each nitrogen. A 0.5 mT peak-to-peak
linewidth was applied. The field-swept spectrum contained 2048 points
from 260 to 360 mT at a microwave frequency of 9.5 GHz and was evaluated
using the matrix method with an orientation grid of {[}19, 4{]}.

This system produces a 72 x 72 spin-Hamiltonian matrix and a
correspondingly dense set of field-dependent transitions. It was
selected to test batched Hamiltonian construction and diagonalization,
resonance-field determination, powder averaging, and interpolation of
numerous transitions onto the experimental field axis.

\subsection*{pepper: high-spin Mn(II)}

The second matrix workload describes an S=5/2 \textsuperscript{55}Mn
center with an isotropic \emph{g} value of 2.0, a 250 MHz isotropic
hyperfine interaction, and zero-field-splitting parameters of D = 600
MHz and E = 100 MHz. A 1 mT linewidth was used, and the spectrum was
calculated over 260--360 mT at 9.5 GHz using 2048 field points and the
same {[}19, 4{]} orientation grid.

Although its 36 x 36 Hamiltonian is smaller than that of the Cu(II)
test, the high electron spin, nuclear spin, and zero-field splitting
generate many transitions and substantial orientation dependence. This
workload therefore tests high-spin Hamiltonian construction and repeated
diagonalization across the powder grid.

\subsection*{pepper: dense-grid perturbative calculation}

The perturbative \emph{pepper} test uses an S = 1/2 system with \emph{g}
= {[}2.0023, 2.0025, 2.0035{]}, one \textsuperscript{14}N nucleus, and
two \textsuperscript{1}H nuclei. Their \emph{A} tensors were {[}20, 20,
90{]}, {[}8, 8, 12{]}, and {[}5, 5, 7{]} MHz, respectively. The
4096-point spectrum covered 330--350 mT at 9.5 GHz, with a 0.2 mT
linewidth and a dense {[}91, 1{]} orientation grid. Resonances were
evaluated using the second-order perturbative method.

This calculation avoids full matrix diagonalization and instead
emphasizes perturbative resonance evaluation, dense orientation
sampling, and accumulation onto a high-resolution field grid. Because
individual calculations complete in substantially less than one second,
their timings are especially sensitive to fixed interpreter, dispatch,
and thread-management overhead.

\subsection*{pepper: strain summation}

The strain workload contains an S = 1/2 \textsuperscript{14}N-coupled
system with \emph{g} = {[}2.008 ,2.006, 2.003{]} and \emph{A} = {[}20,
20, 85{]} MHz. Anisotropic \emph{g}-strain values of {[}0.003, 0.002,
0.001{]} and anisotropic residual broadening specified by {[}10, 10,
30{]} MHz were included together with a 0.3 mT linewidth. The 2048-point
spectrum was calculated from 330 to 350 mT at 9.5 GHz using a {[}61,
4{]} orientation grid.

This test was selected because strain broadening requires the program to
evaluate orientation- and transition-dependent linewidths and to
accumulate many broadened contributions. Its dominant costs therefore
differ from those of a simple matrix calculation and exercise spectral
construction and summation.

\subsection*{pepper: repeated 20-call fitting pattern}

To represent the repeated forward evaluations required during spectral
fitting, a sequence of 20 \emph{pepper} simulations was timed as a
single workload. The model consisted of an S = 1/2 system coupled to
\textsuperscript{14}N, with \emph{A} = {[}20, 20, 85{]} MHz and a 0.3 mT
linewidth. Across the sequence, the first principal \emph{g}-value
(\emph{g\textsubscript{x}}) was varied linearly from 2.000 to 2.010
while the remaining values were held at 2.006 and 2.003. Each spectrum
contained 1024 points over 330-350 mT at 9.5 GHz and used a {[}31, 4{]}
orientation grid.

This repeated-forward benchmark isolates the cumulative cost of
performing the closely related simulations that occur in an iterative
least-squares or parameter-search workflow including repeated setup,
Hamiltonian evaluation, and spectral accumulation.

\subsection*{chili: nitroxide with two coupled nuclei}

The first slow-motion workload represents an S = 1/2 nitroxide coupled
to one \textsuperscript{14}N and one \textsuperscript{1}H nucleus. The
calculation used \emph{g} = {[}2.008,2.006,2.003{]}, nitrogen and proton
hyperfine tensors of {[}20, 20, 85{]} and {[}5, 5, 8{]} MHz,
respectively, and a rotational correlation time of 30 ns. Gaussian and
Lorentzian broadening contributions were each set to 0.1 mT. The
1024-point field sweep covered 330-350 mT at 9.5 GHz, using the default
stochastic-Liouville basis truncation of {[}14, 7, 2, 6{]}.

This workload tests construction of the stochastic Liouville operator,
rotational basis truncation, treatment of multiple coupled nuclei, and
the iterative linear-algebra operations used to obtain a slow-motion
spectrum. These operations have computational characteristics distinct
from the orientation-dependent spin-Hamiltonian diagonalizations used by
\emph{pepper}.

\subsection*{chili: orienting potential}

The second \emph{chili} workload used an S = 1/2,
\textsuperscript{14}N-coupled nitroxide with \emph{g} = {[}2.008, 2.006,
2.003{]}, \emph{A} = {[}20, 20, 85{]} MHz, and a 20 ns rotational
correlation time. The orientational potential was defined by the
coefficient set {[}2, 0, 0, 1.5{]}. Gaussian and Lorentzian linewidth
contributions of 0.1 mT were applied, and the 1024-point spectrum was
calculated over 330--350 mT at 9.5 GHz with the {[}14, 7, 2, 6{]} basis
truncation.

Including an orienting potential introduces coupling between rotational
basis functions and requires explicit treatment of a nonuniform
molecular orientation distribution. This test therefore probes the
additional Liouvillian construction and orientational averaging required
for partially ordered slow-motion systems.

\subsection*{saffron: HYSCORE}

The pulse-EPR workload is a 512 x 512 HYSCORE calculation for an S = 1/2
center coupled to \textsuperscript{14}N and \textsuperscript{1}H. The
\emph{g}-value was 2.0023. \emph{A}-tensors were {[}3, 3, 6{]} MHz for
nitrogen and {[}2, 2, 8{]} MHz for the proton, with a
\textsuperscript{14}N quadrupole tensor of {[}−0.5, −0.5, 1.0{]} MHz.
The calculation used a magnetic field of 350 mT, a 0.1 µs interpulse
delay, a 0.016 µs time increment, and a 91-point orientation grid.

This multidimensional calculation tests orientation-dependent nuclear
Hamiltonians, pulse-sequence evolution, construction of a
two-dimensional time-domain response, and transformation to the HYSCORE
frequency domain. It was included because its memory-access and
tensor-operation patterns differ substantially from those of field-swept
continuous-wave simulations. Both CPU and hybrid CUDA execution were
verified for this workload.

\subsection*{cardamom: stochastic rotational trajectories}

The trajectory workload describes an S = 1/2 nitroxide coupled to
\textsuperscript{14}N, with \emph{g} = {[}2.008, 2.006, 2.003{]},
\emph{A} = {[}20, 20, 85{]} MHz, and a rotational correlation time of 1
ns. The absorption spectrum contained 256 points from 332 to 352 mT at
9.5 GHz. The diffusion model generated 200 rotational trajectories for
each starting orientation, with 1000 propagation steps and spin and
spatial time increments of 10\textsuperscript{-10} s.

This calculation tests stochastic rotational-trajectory generation,
quaternion and rotation-matrix operations, time-dependent
spin-Hamiltonian construction, density-matrix propagation, ensemble
averaging, and conversion of the resulting time-domain response to a
field-domain spectrum. Because \emph{cardamom} does not expose a public
random seed, each timed call generated an independent finite trajectory
ensemble with identical dimensions and simulation parameters. The
benchmark therefore measures the complete trajectory-simulation workflow
while accepting the small run-to-run variation inherent to independent
stochastic ensembles.

\begin{table}[H]
\captionsetup{position=top}
\caption{Matched-host performance measurements used in the manuscript comparison.}
\label{tab:S1}
\centering
\small
\setlength{\tabcolsep}{4pt}
\renewcommand{\arraystretch}{1.3}
\begin{tabular}{@{}>{\raggedright\arraybackslash}p{\dimexpr 0.260000\linewidth-1.666667\tabcolsep\relax}>{\raggedright\arraybackslash}p{\dimexpr 0.180000\linewidth-1.666667\tabcolsep\relax}>{\raggedright\arraybackslash}p{\dimexpr 0.160000\linewidth-1.666667\tabcolsep\relax}>{\raggedright\arraybackslash}p{\dimexpr 0.110000\linewidth-1.666667\tabcolsep\relax}>{\raggedright\arraybackslash}p{\dimexpr 0.170000\linewidth-1.666667\tabcolsep\relax}>{\raggedright\arraybackslash}p{\dimexpr 0.120000\linewidth-1.666667\tabcolsep\relax}@{}}
\toprule
Workload & EasySpin (s) & TorchSpin CPU (s) & ES/TS CPU & TorchSpin CUDA mode (s) & ES/TS CUDA \\
\midrule
pepper: Cu + 2 N, matrix & 0.292 {[}0.229--0.351{]} & 0.847 (32 t) & 0.344 & 0.674 (hybrid) & 0.433 \\
pepper: Mn(II), matrix & 3.03 {[}3.03--3.14{]} & 1.07 (32 t) & 2.83 & 0.954 (hybrid) & 3.18 \\
pepper: perturbative grid & 0.0818 {[}0.072--0.0825{]} & 0.0616 (1 t) & 1.33 & 0.162 (hybrid) & 0.506 \\
pepper: strain summation & 0.814 {[}0.808--0.816{]} & 0.994 (8 t) & 0.82 & 1.23 (hybrid) & 0.661 \\
pepper: 20-call fit pattern & 2.5 {[}2.5--2.5{]} & 1.29 (8 t) & 1.93 & 1.77 (hybrid) & 1.41 \\
chili: nitroxide + proton & 1.15 {[}1.14--1.16{]} & 0.157 (1 t) & 7.32 & N/A & N/A \\
chili: orienting potential & 4.53 {[}4.53--4.53{]} & 3.08 (1 t) & 1.47 & N/A & N/A \\
saffron: HYSCORE & 0.66 {[}0.536--0.667{]} & 0.752 (32 t) & 0.877 & 0.228 (hybrid) & 2.89 \\
cardamom: trajectories & 20.4 {[}20.4--20.4{]} & 9.97 (8 t) & 2.04 & N/A & N/A \\
\bottomrule
\end{tabular}
\end{table}

Table S1 reports the benchmark campaign used for Figure 2 and Figure S5.
\emph{EasySpin}/\emph{TorchSpin} ratios greater than one favor
\emph{TorchSpin}. CUDA-capable calculations are labeled hybrid because
CPU work remained in every measured path; unsupported CPU-only
calculations are reported as not available. CPU and CUDA outputs agreed
to a maximum relative deviation of 3.3 x 10\textsuperscript{-12} and an
NRMSE of 2.2 x 10\textsuperscript{-12} or better.

\FloatBarrier
\section*{Machine-learning experiments}

\subsection*{Cu(II) spectral libraries}

The inverse-model studies used first-derivative X-band Cu(II) powder
spectra spanning a bounded axial-to-rhombic, `blue-copper-like'
parameter domain. Each spectrum represented an S = 1/2 electron coupled
to \textsuperscript{63}Cu and was calculated at 9.5 GHz over 260--360 mT
using 512 field points, D\textsubscript{2h} symmetry, an orientation
grid of {[}31 1{]}, and Gaussian broadening. Seven sampled coordinates
described the centers and rhombicities of the transverse \emph{g} and
hyperfine components, \emph{g\textsubscript{z}},
\emph{A\textsubscript{z}}, and linewidth. The transformation in Table S2
preserves \emph{g\textsubscript{x}} ≤ \emph{g\textsubscript{y}} and
\emph{A\textsubscript{x}} ≤ \emph{A\textsubscript{y}} while sampling
both axial and rhombic cases continuously.

\begin{table}[H]
\captionsetup{position=top}
\caption{Bounded coordinates used to generate the Cu(II) spectral libraries.}
\label{tab:S2}
\centering
\small
\setlength{\tabcolsep}{4pt}
\renewcommand{\arraystretch}{1.3}
\begin{tabular}{@{}>{\raggedright\arraybackslash}p{\dimexpr 0.321435\linewidth-1.333333\tabcolsep\relax}>{\raggedright\arraybackslash}p{\dimexpr 0.287277\linewidth-1.333333\tabcolsep\relax}>{\raggedright\arraybackslash}p{\dimexpr 0.391288\linewidth-1.333333\tabcolsep\relax}@{}}
\toprule
Model coordinate & Sampled interval & Physical transformation \\
\midrule
g\textsubscript{⊥} center & 2.040--2.080 & g\textsubscript{x} = g\textsubscript{⊥} − Δg/2;\par g\textsubscript{y} = g\textsubscript{⊥} + Δg/2 \\
g rhombicity, Δg & 0--0.030 & g\textsubscript{y} − g\textsubscript{x} \\
g\textsubscript{z} & 2.18--2.30 & -- \\
A\textsubscript{⊥} center & 35--85 MHz & A\textsubscript{x} = A\textsubscript{⊥} − ΔA/2;\par A\textsubscript{y} = A\textsubscript{⊥} + ΔA/2 \\
A rhombicity, ΔA & 0--40 MHz & A\textsubscript{y} − A\textsubscript{x} \\
A\textsubscript{z} & 450--650 MHz & -- \\
Gaussian linewidth & 1.4--2.4 mT & -- \\
\bottomrule
\end{tabular}
\end{table}

\subsection*{CNN architecture and supervised training}

All comparisons used the same 1,297,127-parameter residual
one-dimensional CNN. After convolutional encoding, the 128 features at
each of the remaining 32 positions were flattened into a single vector
and passed through fully connected layers containing 256 and 128 units
before producing the seven parameter predictions. During training,
dropout randomly set 5\% of the activations from the first fully
connected layer to zero as a mild regularization step. A sigmoid mapped
each output to its bounded unit coordinate before conversion to the
physical parameters in Table S2.

\subsection*{Differentiable refinement after supervised initialization}

The refinement objective was L = L\textsubscript{derivative} + 0.25
L\textsubscript{integrated} + 0.25 L\textsubscript{weighted}. The
derivative and integrated terms were 1-\emph{C}, where \emph{C} is the
cosine similarity between the corresponding
unit-\emph{L\textsubscript{2}}-normalized spectra. Integration first
removed the derivative mean, formed a cumulative sum, corrected endpoint
drift with a linear ramp, removed the integrated mean, and renormalized
the result. The pointwise squared-error weight was 0.25 + 0.75
\textbar target\textbar{} /max(\textbar target\textbar); the weighted
mean was multiplied by the 512 field points. The parameter-error
coefficient was exactly zero during refinement and in checkpoint
selection. The lowest spectral loss on a fixed set of 128 validation
spectra was selected, with the initial supervised checkpoint included as
epoch zero. No test spectra were removed, and no hyperparameters were
adjusted after test evaluation. Refinement used two RTX 4090 GPUs, with
differentiable spectra assigned alternately to the devices and backward
hooks used to confirm gradient calculations on each GPU.

\begin{table}[H]
\captionsetup{position=top}
\caption{Test-set effects of parameter-label-free spectral refinement after
supervised initialization.}
\label{tab:S3}
\centering
\small
\setlength{\tabcolsep}{4pt}
\renewcommand{\arraystretch}{1.3}
\begin{tabular}{@{}>{\raggedright\arraybackslash}p{\dimexpr 0.437833\linewidth-1.333333\tabcolsep\relax}>{\raggedright\arraybackslash}p{\dimexpr 0.252794\linewidth-1.333333\tabcolsep\relax}>{\raggedright\arraybackslash}p{\dimexpr 0.309372\linewidth-1.333333\tabcolsep\relax}@{}}
\toprule
Metric & Supervised CNN & Physics-refined CNN \\
\midrule
Median clean-spectrum cosine & 0.9768 & 0.9849 \\
Minimum cosine & 0.5934 & 0.8178 \\
Fraction with cosine ≥ 0.950 & 76.6\% & 90.2\% \\
Spectra improved & --- & 373/512 (72.9\%) \\
\bottomrule
\end{tabular}
\end{table}

\begin{table}[H]
\captionsetup{position=top}
\caption{Component-wise parameter MAE before and after parameter-label-free
spectral refinement.}
\label{tab:S4}
\centering
\small
\setlength{\tabcolsep}{4pt}
\renewcommand{\arraystretch}{1.3}
\begin{tabular}{@{}>{\raggedright\arraybackslash}p{\dimexpr 0.335481\linewidth-1.333333\tabcolsep\relax}>{\raggedright\arraybackslash}p{\dimexpr 0.366452\linewidth-1.333333\tabcolsep\relax}>{\raggedright\arraybackslash}p{\dimexpr 0.298067\linewidth-1.333333\tabcolsep\relax}@{}}
\toprule
Parameter & Supervised MAE & Refined MAE \\
\midrule
g\textsubscript{x} & 0.00125 & 0.00116 \\
g\textsubscript{y} & 0.00111 & 0.00099 \\
g\textsubscript{z} & 0.00369 & 0.00396 \\
A\textsubscript{x} (MHz) & 5.72 & 5.89 \\
A\textsubscript{y} (MHz) & 4.25 & 4.44 \\
A\textsubscript{z} (MHz) & 8.26 & 8.42 \\
Linewidth (mT) & 0.0736 & 0.0921 \\
\bottomrule
\end{tabular}
\end{table}

\FloatBarrier
\begin{landscape}
\begin{table}[H]
\captionsetup{position=top}
\caption{Prospective test-set performance for parameter-label-free training from
random initialization.}
\label{tab:S5}
\centering
\footnotesize
\setlength{\tabcolsep}{4pt}
\renewcommand{\arraystretch}{1.3}
\begin{tabular}{@{}>{\raggedright\arraybackslash}p{\dimexpr 0.150000\linewidth-1.800000\tabcolsep\relax}>{\raggedright\arraybackslash}p{\dimexpr 0.100000\linewidth-1.800000\tabcolsep\relax}>{\raggedright\arraybackslash}p{\dimexpr 0.110000\linewidth-1.800000\tabcolsep\relax}>{\raggedright\arraybackslash}p{\dimexpr 0.080000\linewidth-1.800000\tabcolsep\relax}>{\raggedright\arraybackslash}p{\dimexpr 0.080000\linewidth-1.800000\tabcolsep\relax}>{\raggedright\arraybackslash}p{\dimexpr 0.080000\linewidth-1.800000\tabcolsep\relax}>{\raggedright\arraybackslash}p{\dimexpr 0.105000\linewidth-1.800000\tabcolsep\relax}>{\raggedright\arraybackslash}p{\dimexpr 0.105000\linewidth-1.800000\tabcolsep\relax}>{\raggedright\arraybackslash}p{\dimexpr 0.105000\linewidth-1.800000\tabcolsep\relax}>{\raggedright\arraybackslash}p{\dimexpr 0.085000\linewidth-1.800000\tabcolsep\relax}@{}}
\toprule
Model & Median cosine & Fraction ≥ 0.950 & g\textsubscript{x} MAE & g\textsubscript{y} MAE & g\textsubscript{z} MAE & A\textsubscript{x} MAE (MHz) & A\textsubscript{y} MAE (MHz) & A\textsubscript{z} MAE (MHz) & lw MAE (mT) \\
\midrule
Supervised control & 0.9698 & 65.2\% & 0.00144 & 0.00140 & 0.00432 & 6.45 & 4.41 & 10.0 & 0.0855 \\
Label-free seed 1 & 0.9923 & 98.6\% & 0.00118 & 0.000821 & 0.00392 & 6.98 & 5.01 & 9.04 & 0.107 \\
Label-free seed 2 & 0.9921 & 98.6\% & 0.00115 & 0.000879 & 0.00454 & 7.01 & 4.82 & 9.81 & 0.105 \\
Label-free seed 3 & 0.9913 & 98.6\% & 0.00126 & 0.000945 & 0.00444 & 6.96 & 5.05 & 9.87 & 0.124 \\
\bottomrule
\end{tabular}
\end{table}
\end{landscape}

\FloatBarrier
\section*{Cross-engine training-library transfer}

\begin{table}[H]
\captionsetup{position=top}
\caption{Prospective cross-engine comparison averaged across three paired CNN
initialization seeds. Values are mean ± sample standard deviation across
three paired initialization seeds evaluated on the same 512-spectrum
test set; percentage uncertainties are percentage points, not confidence
intervals.}
\label{tab:S6}
\centering
\small
\setlength{\tabcolsep}{4pt}
\renewcommand{\arraystretch}{1.3}
\begin{tabular}{@{}>{\raggedright\arraybackslash}p{\dimexpr 0.248900\linewidth-1.500000\tabcolsep\relax}>{\raggedright\arraybackslash}p{\dimexpr 0.247107\linewidth-1.500000\tabcolsep\relax}>{\raggedright\arraybackslash}p{\dimexpr 0.256887\linewidth-1.500000\tabcolsep\relax}>{\raggedright\arraybackslash}p{\dimexpr 0.247107\linewidth-1.500000\tabcolsep\relax}@{}}
\toprule
Training library & Median cosine & Fraction ≥ 0.950 & Mean NRMSE \\
\midrule
TorchSpin & 0.9656 ± 0.0008 & 61.59 ± 2.07\% & 0.3075 ± 0.0067 \\
EasySpin & 0.9667 ± 0.0004 & 63.93 ± 1.59\% & 0.3005 ± 0.0063 \\
\bottomrule
\end{tabular}
\end{table}

\FloatBarrier
\begin{landscape}
\begin{table}[H]
\captionsetup{position=top}
\caption{Mean component-wise parameter MAE across three paired initialization
seeds in the cross-engine training-library comparison.}
\label{tab:S7}
\centering
\footnotesize
\setlength{\tabcolsep}{4pt}
\renewcommand{\arraystretch}{1.3}
\begin{tabular}{@{}>{\raggedright\arraybackslash}p{\dimexpr 0.180000\linewidth-1.750000\tabcolsep\relax}>{\raggedright\arraybackslash}p{\dimexpr 0.110000\linewidth-1.750000\tabcolsep\relax}>{\raggedright\arraybackslash}p{\dimexpr 0.110000\linewidth-1.750000\tabcolsep\relax}>{\raggedright\arraybackslash}p{\dimexpr 0.110000\linewidth-1.750000\tabcolsep\relax}>{\raggedright\arraybackslash}p{\dimexpr 0.130000\linewidth-1.750000\tabcolsep\relax}>{\raggedright\arraybackslash}p{\dimexpr 0.130000\linewidth-1.750000\tabcolsep\relax}>{\raggedright\arraybackslash}p{\dimexpr 0.130000\linewidth-1.750000\tabcolsep\relax}>{\raggedright\arraybackslash}p{\dimexpr 0.100000\linewidth-1.750000\tabcolsep\relax}@{}}
\toprule
Training library & g\textsubscript{x} & g\textsubscript{y} & g\textsubscript{z} & A\textsubscript{x} (MHz) & A\textsubscript{y} (MHz) & A\textsubscript{z} (MHz) & lw (mT) \\
\midrule
TorchSpin & 0.00142 & 0.00138 & 0.00481 & 5.65 & 4.56 & 10.9 & 0.0799 \\
EasySpin & 0.00141 & 0.00133 & 0.00490 & 5.99 & 4.55 & 10.5 & 0.0807 \\
\bottomrule
\end{tabular}
\end{table}
\end{landscape}

\begin{table}[H]
\captionsetup{position=top}
\caption{CNN inference and TorchSpin reconstruction latency.}
\label{tab:S8}
\centering
\small
\setlength{\tabcolsep}{4pt}
\renewcommand{\arraystretch}{1.3}
\begin{tabular}{@{}>{\raggedright\arraybackslash}p{\dimexpr 0.380428\linewidth-1.500000\tabcolsep\relax}>{\raggedright\arraybackslash}p{\dimexpr 0.213583\linewidth-1.500000\tabcolsep\relax}>{\raggedright\arraybackslash}p{\dimexpr 0.233262\linewidth-1.500000\tabcolsep\relax}>{\raggedright\arraybackslash}p{\dimexpr 0.172727\linewidth-1.500000\tabcolsep\relax}@{}}
\toprule
Operation & Device/configuration & Median {[}IQR{]} (ms) & Per-spectrum value (ms) \\
\midrule
Single CNN inference, resident & CPU, 1 thread & 0.976 {[}0.974--0.985{]} & 0.976 \\
Host normalization + CNN prediction & CPU, 1 thread & 0.994 {[}0.992--0.998{]} & 0.994 \\
Single CNN inference, resident & RTX 4090 & 1.019 {[}1.016--1.024{]} & 1.019 \\
Host normalization + transfer + prediction & RTX 4090 & 1.047 {[}1.045--1.053{]} & 1.047 \\
Batch CNN inference, n = 32 & RTX 4090 & 1.029 {[}1.026--1.034{]} & 0.0321 \\
Batch CNN inference, n = 128 & RTX 4090 & 1.016 {[}1.014--1.021{]} & 0.00794 \\
Batch CNN inference, n = 512 & RTX 4090 & 1.266 {[}1.264--1.272{]} & 0.00247 \\
TorchSpin forward reconstruction & RTX 4090 & 68.42 {[}68.37--68.50{]} & 68.42 \\
TorchSpin forward reconstruction & CPU, 1 thread & 105.73 {[}105.55--107.96{]} & 105.73 \\
\bottomrule
\end{tabular}
\end{table}


\begingroup
\small
\begin{thebibliography}{99}
\bibitem{ref1}
S. Stoll, A. Schweiger, EasySpin, a comprehensive software package for
spectral simulation and analysis in EPR, \emph{J. Magn. Reson.} 178
(\textbf{2006}) 42--55. https://doi.org/10.1016/j.jmr.2005.08.013.

\bibitem{ref2}
S. Stoll, R.D. Britt, General and efficient simulation of pulse EPR
spectra, \emph{Phys. Chem. Chem. Phys.} 11 (\textbf{2009}) 6614.
https://doi.org/10.1039/b907277b.

\bibitem{ref3}
D.T. Petasis, M.P. Hendrich, Quantitative Interpretation of
Multifrequency Multimode EPR Spectra of Metal Containing Proteins,
Enzymes, and Biomimetic Complexes, \emph{Methods Enzymol.} 563
(\textbf{2015}) 171--208. https://doi.org/10.1016/bs.mie.2015.06.025.

\bibitem{ref4}
H.J. Hogben, M. Krzystyniak, G.T.P. Charnock, P.J. Hore, I. Kuprov,
Spinach -- A software library for simulation of spin dynamics in large
spin systems, \emph{J. of Magn. Reson.} 208 (\textbf{2011}) 179--194.
https://doi.org/10.1016/j.jmr.2010.11.008.

\bibitem{ref5}
G.R. Hanson, K.E. Gates, C.J. Noble, M. Griffin, A. Mitchell, S. Benson,
XSophe-Sophe-XeprView ® . A computer simulation software suite (v.
1.1.3) for the analysis of continuous wave EPR spectra, \emph{J. Inorg.
Biochem.} 98 (\textbf{2004}) 903--916.
https://doi.org/10.1016/j.jinorgbio.2004.02.003.

\bibitem{ref6}
A.T. Taguchi, E.D. Evans, S.A. Dikanov, R.G. Griffin, Convolutional
Neural Network Analysis of Two-Dimensional Hyperfine Sublevel
Correlation Electron Paramagnetic Resonance Spectra, \emph{J. Phys.
Chem. Lett.} 10 (\textbf{2019}) 1115--1119.
https://doi.org/10.1021/acs.jpclett.8b03797.

\bibitem{ref7}
S.G. Worswick, J.A. Spencer, G. Jeschke, I. Kuprov, Deep neural network
processing of DEER data, Science Advances 4 (\textbf{2018}) eaat5218.
https://doi.org/10.1126/sciadv.aat5218.

\bibitem{ref8}
J. Keeley, T. Choudhury, L. Galazzo, E. Bordignon, A. Feintuch, D.
Goldfarb, H. Russell, M.J. Taylor, J.E. Lovett, A. Eggeling, L. Fábregas
Ibáñez, K. Keller, M. Yulikov, G. Jeschke, I. Kuprov, Neural networks in
pulsed dipolar spectroscopy: A practical guide, \emph{J. Magn. Reson.}
338 (\textbf{2022}) 107186. https://doi.org/10.1016/j.jmr.2022.107186.

\bibitem{ref9}
R. Newbury, J. Collins, K. He, J. Pan, I. Posner, D. Howard, A. Cosgun,
A Review of Differentiable Simulators, \emph{IEEE Access} 12 (\textbf{2024)}
97581--97604. https://doi.org/10.1109/ACCESS.2024.3425448.

\bibitem{ref10}
S.S. Schoenholz, E.D. Cubuk, JAX, M.D.: A Framework for Differentiable
Physics, (\textbf{2019}). https://doi.org/10.48550/ARXIV.1912.04232.

\bibitem{ref11}
J.G. Greener, Differentiable simulation to develop molecular dynamics
force fields for disordered proteins, \emph{Chem. Sci.} 15
(\textbf{2024}) 4897--4909. https://doi.org/10.1039/D3SC05230C.

\bibitem{ref12}
A. Paszke, S. Gross, F. Massa, A. Lerer, J. Bradbury, G. Chanan, T.
Killeen, Z. Lin, N. Gimelshein, L. Antiga, A. Desmaison, A. Köpf, E.
Yang, Z. DeVito, M. Raison, A. Tejani, S. Chilamkurthy, B. Steiner, L.
Fang, J. Bai, S. Chintala, PyTorch: An Imperative Style,
High-Performance Deep Learning Library, (\textbf{2019}).
https://doi.org/10.48550/ARXIV.1912.01703.

\bibitem{ref13}
B.L. Bales, EPR of frozen aqueous solutions of Fremy's salt: Resolution
enhancement of polycrystalline spectra, \emph{Chem. Phys. Lett.} 10
(\textbf{1971}) 361--364. https://doi.org/10.1016/0009-2614(71)80309-3.

\bibitem{ref14}
J.H. Freed, Theory of Slow Tumbling ESR Spectra for Nitroxides, in:
\emph{Spin Labeling}, Elsevier, \textbf{1976}: pp. 53--132.
https://doi.org/10.1016/B978-0-12-092350-2.50008-4.

\bibitem{ref15}
E. Etienne, A. Pierro, K.C. Tamburrini, A. Bonucci, E. Mileo, M.
Martinho, V. Belle, Guidelines for the Simulations of Nitroxide X-Band
cw EPR Spectra from Site-Directed Spin Labeling Experiments Using
SimLabel, \emph{Molecules} 28 (\textbf{2023}) 1348.
https://doi.org/10.3390/molecules28031348.

\bibitem{ref16}
L.G. Rowan, E.L. Hahn, W.B. Mims, Electron-Spin-Echo Envelope
Modulation, \emph{Phys. Rev.} 137 (\textbf{1965}) A61--A71.
https://doi.org/10.1103/PhysRev.137.A61.

\bibitem{ref17}
E.J. Reijerse, S.A. Dikanov, Electron spin echo envelope modulation
spectroscopy on orientationally disordered systems: Line shape
singularities in \emph{S} =1/2, \emph{I} =1/2 spin systems, \emph{J. of
Chem. Phys.} 95 (\textbf{1991}) 836--845.
https://doi.org/10.1063/1.461091.

\bibitem{ref18}
L. Ardizzone, J. Kruse, S. Wirkert, D. Rahner, E.W. Pellegrini, R.S.
Klessen, L. Maier-Hein, C. Rother, U. Köthe, Analyzing Inverse Problems
with Invertible Neural Networks, (\textbf{2018}).
https://doi.org/10.48550/ARXIV.1808.04730.

\bibitem{ref19}
B.L. Vallee, R.J. Williams, Metalloenzymes: the entatic nature of their
active sites., \emph{Proc. Natl. Acad. Sci. U.S.A.} 59 (\textbf{1968})
498--505. https://doi.org/10.1073/pnas.59.2.498.

\bibitem{ref20}
J.C. Cox, B.G. Malmström, EPR studies on the blue copper protein,
rusticyanin: A protein involved in Fe\textsuperscript{2+} oxidation at
pH 2.0 in \emph{Thiobacillus ferro‐oxidans}, \emph{FEBS Lett.} 93
(\textbf{1978}) 157--160. https://doi.org/10.1016/0014-5793(78)80826-6.

\bibitem{ref21}
E.I. Solomon, R.K. Szilagyi, S. DeBeer George, L. Basumallick,
Electronic Structures of Metal Sites in Proteins and Models: 
Contributions to Function in Blue Copper Proteins, \emph{Chem. Rev.} 104
(\textbf{2004}) 419--458. https://doi.org/10.1021/cr0206317.
\end{thebibliography}

\endgroup
\end{document}